\documentclass[fleqn,usenatbib]{mnras}

\usepackage{newtxtext,newtxmath}

\usepackage[T1]{fontenc}

\DeclareRobustCommand{\VAN}[3]{#2}
\let\VANthebibliography\thebibliography
\def\thebibliography{\DeclareRobustCommand{\VAN}[3]{##3}\VANthebibliography}

\usepackage{graphicx}	
\usepackage{amsmath}	
\usepackage{caption, subcaption} 
\usepackage{multirow}
\usepackage{orcidlink}

\defcitealias{Walmsley2021}{W22} 
\defcitealias{Walmsley18}{W19} 
\defcitealias{DominguezSanchez2023}{DS23} 
\defcitealias{Desmons2023DetectingLearning}{DBL24} 

\title[Uncovering Tidal Treasures]{Uncovering Tidal Treasures: Automated Classification of Faint Tidal Features in DECaLS Data}

\author[A. J. Gordon et al.]{
Alexander J. Gordon,\thanks{E-mail: alexander.gordon@ed.ac.uk}\orcidlink{0009-0006-4035-5019}
Annette M. N. Ferguson,
and Robert G. Mann
\\
Institute for Astronomy, University of Edinburgh, Royal Observatory, Blackford Hill, Edinburgh EH9 3HJ, UK}

\date{Accepted XXX. Received YYY; in original form ZZZ}

\pubyear{2024}

\begin{document}
\label{firstpage}
\pagerange{\pageref{firstpage}--\pageref{lastpage}}
\maketitle

\begin{abstract}
Tidal features are a key observable prediction of the hierarchical model of galaxy formation and contain a wealth of information about the properties and history of a galaxy. Modern wide-field surveys such as LSST and Euclid will revolutionise the study of tidal features. However, the volume of data will prohibit visual inspection to identify features, thereby motivating a need to develop automated detection methods. This paper presents a visual classification of $\sim2,000$ galaxies from the DECaLS survey into different tidal feature categories: \textit{arms}, \textit{streams}, \textit{shells}, and \textit{diffuse}. We trained a Convolutional Neural Network (CNN) to reproduce the assigned visual classifications using these labels. Evaluated on a testing set where galaxies with tidal features were outnumbered $\sim1:10$, our network performed very well and retrieved a median $98.7\pm0.3$, $99.1\pm0.5$, $97.0\pm0.8$, and $99.4^{+0.2}_{-0.6}$ per cent of the actual instances of \textit{arm}, \textit{stream}, \textit{shell}, and \textit{diffuse} features respectively for just 20 per cent contamination. A modified version that identified galaxies with any feature against those without achieved scores of $0.981^{+0.001}_{-0.003}$, $0.834^{+0.014}_{-0.026}$, $0.974^{+0.008}_{-0.004}$, and $0.900^{+0.073}_{-0.015}$ for the accuracy, precision, recall, and F1 metrics, respectively. We used a Gradient-weighted Class Activation Mapping analysis to highlight important regions on images for a given classification to verify the network was classifying the galaxies correctly. This is the first demonstration of using CNNs to classify tidal features into sub-categories, and it will pave the way for the identification of different categories of tidal features in the vast samples of galaxies that forthcoming wide-field surveys will deliver.
\end{abstract}

\begin{keywords}
galaxies: evolution -- galaxies: formation -- galaxies: interactions -- galaxies: structure -- methods: observational -- methods: statistical
\end{keywords}




\section{Introduction}\label{sec:intro}

Faint tidal features are a crucial observational tracer of hierarchical galaxy formation and evolution in the $\Lambda$CDM Universe. In this framework, galaxies grow in size and mass by accreting or merging with other galaxies \citep[e.g.][]{White1978CoreClustering, White1991GalaxyClustering}, these mergers are often described as being major (mass ratio of $\sim0.1$ -- 1) or minor \citep[mass ratio $\lesssim0.1$, e.g.][]{Davies2015, Hendel2015TidalGalaxies}. A significant proportion of a galaxy's mass can originate from material accreted from other galaxies and, in particular, from minor mergers \citep[e.g.][]{Oser2010TheFormation, Ownsworth2014MinorTechniques} which are expected to be more common than major mergers \citep[e.g.][]{Fakhouri2010}. Tidal features are the debris left behind by these interactions \citep[e.g.][]{Toomre1972GalacticTails} either from recently accreted material from minor mergers or late-stage relics of major ones. The morphology of these features varies widely \citep[see, e.g.][]{Quinn1982TheMergers}, however most are typically low surface brightness (LSB). This LSB nature of tidal features makes them very challenging to study, and shallow imaging surveys will generally miss all but the most extreme examples. However, tidal features in galaxy outskirts are generally long-lived \citep[$\sim 1 - 5\text{Gyr}$; see, e.g.][]{2014A&A...566A..97J, Mancillas2019ProbingSubstructures, 2020ApJ...905..154Y} and predicted to be increasingly common at decreasing surface brightness levels \citep[e.g.][]{Johnston2008, Casanova21}.

Tidal features contain a wealth of information about a galaxy’s past. As tidal features are direct byproducts of galaxy mergers, it is possible to reconstruct the merger history of a galaxy by studying its tidal features \citep[e.g.][]{Johnston1999CONSTRAININGDEBRIS, Johnston2008}. In particular, the morphology of the tidal features can be connected to the stellar kinematics and formation history of the galaxy \citep[e.g.][]{Valenzuela2022AHistory} and can probe the orbital distribution of progenitor galaxies \citep[e.g.][]{Hendel2015TidalGalaxies}. Hence, there is interest in measuring the morphology of the tidal features, over and above generating a large sample where some kind of feature is present. Mergers and interactions have significant impacts on the stellar kinematics of galaxies \citep[e.g.][]{Yoon2022EvidenceGalaxies}, and tidal stripping could be a considerable source of star formation suppression \citep[e.g.][]{Spilker2022StarGalaxy}. Tidal features can also be studied to give deeper insights into dark matter \citep[e.g.][]{Sanderson2015Action-spacePotential, Bovy2016THESTREAMS, Pearson2022MappingA} and as tests of the cosmological model \citep[e.g.][]{Johnston2001, Conselice2014GalaxyParameters}. However, to better understand this, it is necessary to build up a significant sample of identified tidal features, which requires probing deep limiting surface brightnesses.

Forthcoming wide-field surveys, such as the Legacy Survey of Space and Time \citep[LSST;][]{Ivezic2019LSST:Products} at the Vera C. Rubin Observatory and the European Space Agency’s Euclid \citep{Laureijs2011Euclid:Universe} , as well as other deep galaxy imaging projects \citep[e.g. ARRAKIHS;][]{ARRAKIHS}, will revolutionise the study of the low surface brightness Universe, including tidal features. These surveys will probe deep limiting surface brightnesses over a vast sky area and uncover many galaxies with tidal features. For example, Euclid will image around 15000 deg$^2$ to a limit of $\mu_{\text{VIS}}=29.5$ mag arcsec$^{-2}$ \citep{Borlaff2022EuclidEuclid/VIS} and LSST almost 18000 deg$^2$ to $\mu_{r}=30.3$ mag arcsec$^{-2}$ after 10 years \citep{Yoachim2022SMTN-016:Derivations}. LSST alone is predicted to discover millions of tidal features \citep{Martin2022PreparingImages}. However, the scientific potential of this data will only be realised if methods are developed to manage the vast volume of data.

Most of the effort to classify or characterise tidal features has involved one or more experts spending extensive amounts of time visually inspecting each image of a galaxy to identify whether or not a tidal feature is present \citep[see e.g.][]{Atkinson13, Bilek2020CensusSurvey, Martin2022PreparingImages, Sola2022CharacterizationImages, Desmons2023GalaxySamples}, even where there were some automated aspects to the process \citep[e.g.][]{Kado-Fong2018TidalChannels}. This is all very well for a modest number of samples; however, this process will only scale to a small volume of data from forthcoming surveys. It is, therefore, necessary to develop an automated process to identify and classify tidally-disturbed galaxies.

This issue of too much data is not isolated to tidal feature detection. Many researchers are turning to machine learning (ML) to automate and accelerate analysis with data from these modern surveys. There has been a significant amount of effort to use ML to automate the classification of the overall morphology of a galaxy, such as separating between spirals and ellipticals \citep[see e.g.][]{2015MNRAS.450.1441D, 2015ApJS..221....8H, 2018MNRAS.476.3661D, Gonzalez2018GalaxyAugmentation, Barchi2020MachineStudy, 2020MNRAS.491.1554W, Fielding2021AClassification, Reza2021GalaxyLearning, 2021MNRAS.506.1927V, Zhang2022ClassifyingLearning, Xu2023FromAdaptation}. Several different ML approaches can work for this task; however, \citet{Cheng2020OptimizingImaging} demonstrated that Convolutional Neural Networks (CNNs) were often the best performing. CNNs are a popular and state-of-the-art method in computer vision problems and perform remarkably well at image classification tasks; because of this, they have become widely used in astronomy \citep[see, e.g.][for a review]{Fluke2020SurveyingAstronomy, 2023PASA...40....1H}. Some researchers have attempted to use unsupervised ML to address the scalability issue \citep[e.g.][]{Hocking2018AnLearning, Martin2020GalaxyLearning, Spindler_20, Fielding2022TheTechniques}, which has the benefit of not requiring pre-labelled training data and hence much less effort by inspectors. However, it is not always straightforward or guaranteed that the representation clusters will match astrophysical phenomena. 

In a related task to identifying tidal features, some researchers have attempted different ML approaches to classify instances of galaxy mergers with several using Neural Networks or CNNs \citep[see, e.g.][]{Ackermann2018UsingMergers, Pearson2019IdentifyingLearning, Ciprijanovic2020DeepMerge:Networks, Ferreira2020GalaxyData, 2023A&A...669A.141S}. Most of these approaches obtained similar, if not better, results than more traditional numerical markers such as concentration, asymmetry, or a combination of markers \citep[see, e.g.][for an example of a more conventional approach]{Nevin2019AccurateImaging}. Furthermore, ML has been used to identify strong gravitational lenses in images \citep[e.g.][]{Jacobs2017FindingNetworks, Petrillo2017FindingNetworks, Lanusse2018CMUFinding, Petrillo2019LinKS:Networks}, which is a similar problem to that of tidal feature detection with extended regions of low surface brightness material. Thus, ML and CNNs are highly useful -- and, arguably, essential -- tools for researchers to extract meaningful science with the volume of data from modern surveys and are likely suitable tools to classify tidal features.

There have been some attempts in the literature to perform binary classification of tidal features using supervised ML \citep[e.g.][]{Walmsley18, DominguezSanchez2023, Desmons2023DetectingLearning}. A binary classifier aims only to decide whether or not a tidal feature is present without then attempting to categorise those features. While binary classifications can indicate the frequency of minor mergers and accretions, insights into galaxy assembly come from more detailed analyses, such as the nature of the different features and their exact morphologies \citep[e.g.][]{Varghese2011StellarReconstruction, Hendel2015TidalGalaxies, Nibauer2023ConstrainingStreams}. 

A particularly relevant study is that of \citet[][hereinafter \citetalias{Walmsley18}]{Walmsley18} who trained three binary classifiers to detect tidal features in Canada-France-Hawaii Telescope Legacy Survey \citep[CFHTLS;][]{Gwyn2012TheCatalogs} data. This data covered around 170 deg$^2$ to a depth of $\mu_r \sim 27.1$ mag arcsec$^{-2}$ and the galaxy sample had previously been visually inspected by \citet{Atkinson13} for the presence of tidal features. \citetalias{Walmsley18} used those labels to train their classifiers, with just 305 galaxies showing evidence of a tidal feature. The first of their classifiers was a single CNN, with the other two being five single classifiers combined into an ensemble with different configurations. They found that the performance was better for the ensembles than a single CNN and was similar for both configurations. The ensembles recovered an average of $76\pm2$ per cent of the true instances of galaxies with tidal features (true positive rate) for a contamination of only 22 per cent. Contamination here refers to galaxies without tidal features classified as having tidal features (false positive rate).

Similarly \citet[][hereinafter \citetalias{DominguezSanchez2023}]{DominguezSanchez2023} used 5,835 mock images of Hyper Suprime-Cam Subaru Strategic Program \citep[HSC-SSP;][]{Aihara2018TheDesign} galaxies, generated from \textsc{NewHorizon} \citep{2021A&A...651A.109D}. These mock images were previously visually inspected by \citet{Martin2022PreparingImages}. The images were generated from just 36 parent galaxies with noise added to simulate different limiting surface brightnesses and then separated into two training sets: the original sample of $\mu_r = 28-35$ mag arcsec$^{-2}$ and a shallower set with additional low depth images, $\mu_r = 26-35$ mag arcsec$^{-2}$. These two training sets were then used to train CNNs with the deeper dataset, which we estimated by reading their figure to have a TPR of $0.875\pm0.005$, performing better than the shallower set with TPR = $0.815\pm0.005$, at a level of 20 per cent contamination. The authors attempted to transfer their network to real HSC-SSP data but reported that the results were significantly degraded compared to their simulated counterparts, and no performance statistics were provided.

\citet[][hereinafter \citetalias{Desmons2023DetectingLearning}]{Desmons2023DetectingLearning} used HSC-SSP Ultra-Deep data, covering only 3.5 deg$^2$ of the sky but with a much greater surface brightness limit of $\mu_r \sim 29.82$ mag arcsec$^{-2}$, to train a binary classifier. The sample was also small at just 380 galaxies with tidal features. Still, it was significantly improved in depth for low surface brightness detection. Instead of a purely supervised or unsupervised approach, \citetalias{Desmons2023DetectingLearning} used a self-supervised process that is somewhat of a hybrid between supervised and unsupervised. Overall this allowed the network to be trained with fewer labelled data, saving inspectors significant time and effort. \citetalias{Desmons2023DetectingLearning} trained their network to identify important parts of the images using an unsupervised approach that compared augmented -- e.g. rotated, translated, etc -- versions of the same image. They then trained the final part of the network, which identified whether or not a tidal feature was present, using the already trained part to extract important image features and a small number of labelled data. \citetalias{Desmons2023DetectingLearning} report that their network achieved a TPR of $0.94\pm0.1$ for $\text{FPR}=0.2$.

In this paper, we go beyond simply the detection of tidal features in galaxies and conduct the first exploration of Convolutional Neural Networks to classify tidal features into appropriate sub-categories. Section \ref{sec:data} details the data and the selection criteria we applied to produce a training set. Section \ref{sec:visual_classification} discusses how we used visual inspection to generate a label for each galaxy representing its tidal features. Section \ref{sec:CNNs} presents the network we used and the results of applying that to recreate our labels. Section \ref{sec:discussion} discusses some suggested improvements and the issues we encountered, and we conclude in Section \ref{sec:conclusions}.

\section{The Data}\label{sec:data}
To train a CNN to achieve a satisfactory level of accuracy, we needed an extensive dataset comprising identified and labelled tidal features. Despite the depth being notably shallower than the expectations for forthcoming imaging surveys, we chose to use data from the Dark Energy Camera Legacy Survey \citep[DECaLS;][]{Dey2019OverviewSurveys}. This was motivated by the fact that it is one of the largest uniform datasets currently available, and a significant portion of it has previously undergone morphological study by \citet[hereinafter \citetalias{Walmsley2021}]{Walmsley2021}, which we could build upon for our work on tidal feature identification. This prior work provided an initial indication of the presence of a tidal disturbance, enabling us to prune down the sample to contain only those sources most deserving of visual inspection. Furthermore, this dataset served as an excellent testbed for developing our concepts and allowed us to compare with previous works on the subject.

\subsection{DECaLS}

\begin{figure}
    \includegraphics[width=\columnwidth]{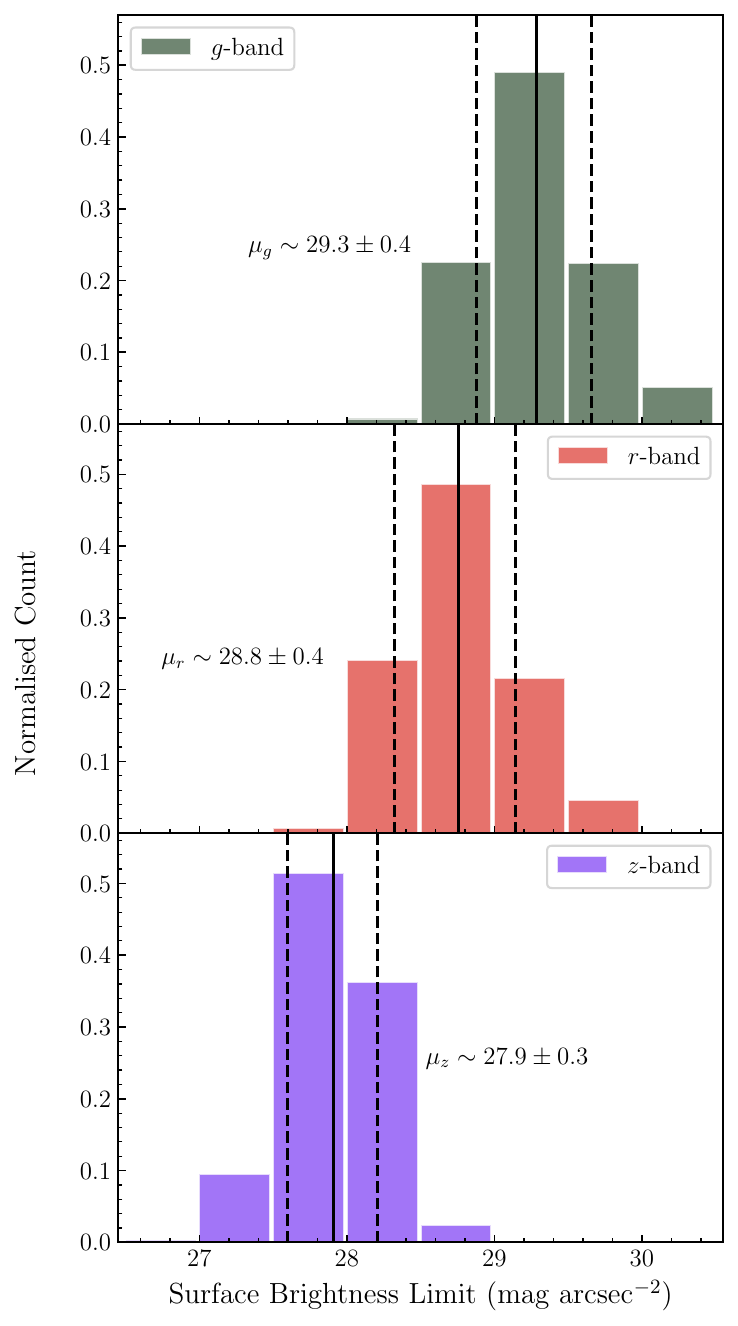}
    \caption{Histogram of the $3\sigma$ limiting surface brightness magnitudes for DECaLS images in the $g$ (top), $r$ (middle), and $z$ (bottom) bands. The median (solid line) and 68.2 per cent confidence interval (dashed lines) were estimated by determining the limit in $\sim$2,000 FITS cutout images of galaxies. For each image, we followed the method set out in \citet{Roman2020GalacticImaging} and used randomly sampled $10 \times 10$ arcsec$^2$ boxes.}
    \label{fig:surface_brightness_limit}
\end{figure}

DECaLS aimed to identify targets for the Dark Energy Spectroscopic Instrument (DESI) survey and used the 4m Blanco telescope at the Cerro Tololo Inter-American Observatory in Chile. Around 9000 deg$^2$ of the sky was imaged using a three-pass tiling system, where each pass was slightly offset from the others. The survey was kept as uniform as possible by using dynamic observation to automatically select targets and exposure times based on observing conditions. The result of this was around 835 million objects imaged in the \textit{g}, \textit{r}, and \textit{z} bands with a native pixel scale of 0.262 arcsec px$^{-1}$. The limiting point source depth in the \textit{g} band is 23.95 (AB) for a median 5$\sigma$ detection. We estimated the 3$\sigma$ limiting surface brightness magnitude in $10\arcsec \times 10\arcsec$ boxes to be $\mu_r \sim 28.8$ mag arcsec$^{-2}$, by following the procedure set out in \citet{Roman2020GalacticImaging}. Figure \ref{fig:surface_brightness_limit} presents normalised histograms and median estimates of the limiting surface brightness magnitude across $\sim2000$ images of galaxies for each DECaLS band. The images used were FITS cutouts of galaxies downloaded from the Legacy Survey\footnote{\url{https://www.legacysurvey.org/}} cutout service\footnote{\url{https://www.legacysurvey.org/viewer/urls}} at the native pixel scale. Our estimates of the DECaLS surface brightness limiting depth are similar to those quoted in other studies \citep[e.g.][]{2021A&A...656A..44R, 2021A&A...652A..48M}.

\subsection{Galaxy Zoo DECaLS}\label{sec:gzd_work}
We made use of publicly available data products\footnote{\url{https://zenodo.org/record/4573248}} produced by \citetalias{Walmsley2021} in the form of both galaxy images and classifications. In \citetalias{Walmsley2021}, the authors used DECaLS data to construct RGB Portable Network Graphics (PNG) images of the subset of galaxies that were included in the NASA Sloan Atlas (NSA)\footnote{\url{http://nsatlas.org/}}. The NSA contains a catalogue of various parameters for local galaxies which were primarily imaged in SDSS and GALEX. Basing the sample on the NSA introduced two selection cuts: most selected galaxies are brighter than $m_r = 17.77$ unless they were included in deeper SDSS fields, and the sample has a maximum redshift of $z = 0.15$. Additionally, \citetalias{Walmsley2021} added two further cuts limiting the selection to galaxies with a Petrosian radius of at least 3 arcsec -- such that the galaxies were sufficiently resolved for classification -- and discarding any incomplete images where more than 20 per cent of the pixels in any band were missing. In total, the remaining sample comprised almost 314,000 galaxies.

\citetalias{Walmsley2021} constructed their PNG images to be $424\times424$ pixels by downloading the FITS files from the Legacy Survey cutout service. They ensured that the whole galaxy was suitably visible on the image by resizing the image to an appropriate interpolated arcsec per pixel scale. They then multiplied the \textit{g}, \textit{r}, and \textit{z} band fluxes by 125.0, 71.43, and 52.63, respectively, such that the false colour images had an appropriate range of colours in RGB; \citetalias{Walmsley2021} chose these values by hand. The pixels with low fluxes were then desaturated to avoid a speckled effect in the images. Finally, the fluxes were scaled by $\sinh^{-1}(x)$ to compensate for the wide range of pixel values, linearly rescaled to be in the range $(-0.5, 300)$ to remove the brightest pixels, and then clipped to the usual $(0, 255)$ range for PNG files.

\begin{figure}
    \begin{subfigure}[b]{\columnwidth}
        \includegraphics[width=\columnwidth]{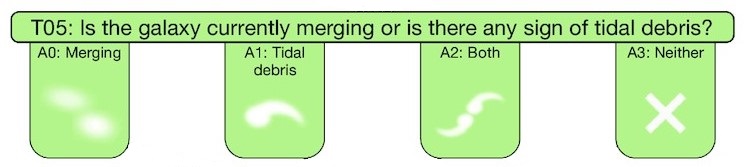}
        \caption{Merger question for DR1\&2 campaigns.}
        \label{fig:walmsley22_tree-subfig:old}
    \end{subfigure}

    \begin{subfigure}[b]{0.6\columnwidth}
        \includegraphics[width=\columnwidth]{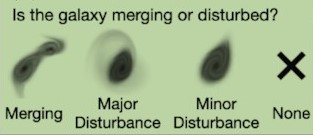}
        \caption{Merger question for DR5 campaigns.}
        \label{fig:walmsley22_tree-subfig:new}
    \end{subfigure}
    
    \caption{Galaxy Zoo DECaLS decision tree merger question presented to volunteers. The question aimed to establish if galaxies were merging or disturbed in some manner and changed between the DR1\&2 (\subref{fig:walmsley22_tree-subfig:old}) and DR5 campaigns (\subref{fig:walmsley22_tree-subfig:new}). Both figures were adapted from \citet[\citetalias{Walmsley2021}]{Walmsley2021}.}
    \label{fig:walmsley22_tree}
\end{figure}

\citetalias{Walmsley2021} used three \textit{Galaxy Zoo: DECaLS} (hereinafter \textit{GZD}) campaigns to obtain volunteer classifications for all galaxy images; each campaign corresponded to a different DECaLS data release -- DR1, DR2, and DR5. Volunteers were asked a series of questions in a decision-tree structure to classify the bulk morphology of the galaxy, such as identifying if the galaxy was early- or late-type, how many spiral arms there were, and if a bar or bulge was present. Between DR1\&2 and DR5, the decision tree questions were changed to improve clarity and direct classifications towards more specific scientific goals. For the DR1\&2 campaigns, a median of 38 volunteers responded for each galaxy. However, the significantly larger DR5 campaign had a median of just five volunteers per galaxy.

As part of both decision trees, the volunteers were asked to indicate if the galaxy was merging or disturbed in some manner, the latter of which we took as a proxy for the potential to have a tidal feature. Figure \ref{fig:walmsley22_tree} provides the options for indicating a disturbance presented to volunteers in both the DR1\&2 (\subref{fig:walmsley22_tree-subfig:old}) and DR5 (\subref{fig:walmsley22_tree-subfig:new}) decision trees. \citetalias{Walmsley2021} changed the possible answers for the merger question between DR1\&2 and DR5 to reflect what the volunteers would see in the image more directly. In particular, the major and minor disturbance labels referred to the extent and size of the debris, not necessarily the origin of the disturbance as a major or minor event. Hence, galaxies indicated as either a major or a minor disturbance were likely to have tidal features.

The classifications from the \textit{GZD} campaigns were then processed and used to train a machine-learning classifier. The classifier constructed by \citetalias{Walmsley2021} was a Bayesian Convolutional Neural Network based on the \textsc{EfficientNetB0} architecture \citep{EfficientNet}. It was trained to predict how volunteers would have responded to the DR5 decision tree, and its accuracy varied from 77 to 99 per cent compared to the volunteer responses depending on the question being considered. We used the catalogue of predictions\footnote{\url{https://zenodo.org/records/4573248/files/gz_decals_auto_posteriors.csv?download=1}} and the corresponding images as a starting point for our analysis.

\subsection{Sample Selection}\label{sec:sample_selection}
From the \textit{GZD} data, we generated a sample of galaxies with a high likelihood of having tidal features by imposing three selection criteria in addition to the cuts introduced in \citetalias{Walmsley2021}. We first limited the sample to have an absolute magnitude in the range $-19 \geq M_r \geq -22$. Removing the faintest galaxies restricted the number of contaminating intrinsically irregular galaxies, most of which are fainter than $M_r \geq -19$ in the local universe \citep{Ann2015AUniverse}. At the surface brightness depth of DECaLS, faint irregular galaxies were often difficult to distinguish from tidally-disturbed galaxies. We removed these galaxies to avoid generating necessarily uncertain classification labels for them.

The remaining two selection criteria were based on the machine-learning predictions for questions in the catalogue. Images predicted to have a greater than 0.1 chance of having an image artefact were removed. Finally, the last criterion selected the galaxies most likely to have some tidal feature, allowing us to reduce the number of galaxies to a manageable amount. Any galaxy with a prediction of greater than 0.4 in either the \verb|merging_major-disturbance_fraction| (hereinafter \verb|major|) or \verb|merging_minor-disturbance_fraction| (hereinafter \verb|minor|) columns in the catalogue was assumed to potentially include a tidal feature. \citetalias{Walmsley2021} recommended using \verb|major| greater than 0.6 and \verb|minor| greater than 0.4 to identify post-merger and low surface brightness galaxies. We randomly sampled and inspected a small number of galaxies with varying values of both \verb|major| and \verb|minor|, and from this extended our threshold to include both \verb|major| and \verb|minor| values greater than 0.4.

\begin{table}
    \caption{Number of galaxies that passed each selection criterion. We applied three selection criteria to the \citetalias{Walmsley2021} sample of 313,789 galaxies. The absolute magnitude of the galaxies was limited to the range $-19 \geq M_r \geq -22$, and images with a prediction greater than 0.1 of an artefact were eliminated. The sample was separated into those potentially having a tidal feature and those assumed not to by considering the output of the \citetalias{Walmsley2021} classifier for the merger question. See the text for full details on this process. Galaxies where the images were unavailable were removed. Finally, some galaxies were removed after the visual inspection (see Section \ref{sec:inspection_results}).}
    \centering
    \label{tab:selection_criteria}
    \begin{tabular}{l c c}
        \hline
        \multirow{2}{*}{Criterion} & \multicolumn{2}{c}{Number of galaxies} \\
        & Tidal Feature & No Tidal Feature \\
        \hline
         & \multicolumn{2}{c}{313,789}\\
        Magnitude cut & \multicolumn{2}{c}{$\downarrow$} \\
         & \multicolumn{2}{c}{255,337}\\
        Artefact cut & \multicolumn{2}{c}{$\downarrow$} \\
         & \multicolumn{2}{c}{198,473}\\
        Disturbance cut & \multicolumn{2}{c}{$\swarrow \quad \searrow$}\\
        \multirow{3}{7em}{Remove missing images} & 1,935 & 10,123\\
         & $\downarrow$ & $\downarrow$ \\
         & 1,928 & 9,981 (91.3\%) \\
        Post-inspection & $\downarrow$ & \\
         & 956 (8.7\%) & \\
        \hline
    \end{tabular}
\end{table}

Of the roughly 314,000 galaxies, 1,935 passed the above selection criteria and were assumed to be tidally disturbed. Images for seven of these were not included in the data set produced by \citetalias{Walmsley2021}, giving a set of 1,928 galaxies to visually inspect for tidal features. Furthermore, we generated a complementary sample of galaxies assumed not to have tidal features by randomly sampling those with \verb|minor|, \verb|major|, and \verb|merging_merger_fraction| predictions of less than 0.08. The breakdown of how many galaxies passed each criterion and were subsequently separated into assumed to have a tidal feature and assumed not is provided in Table \ref{tab:selection_criteria}. We did not visually inspect this undisturbed sample, but it was included in the training set for the convolutional neural network.

\section{Initial Visual Inspection}\label{sec:visual_classification}

\subsection{Preparation for Inspection}\label{sec:prep_for_inspect}
We began by regenerating the thumbnails of the tidal feature sample for visual inspection, using a stretch that enhanced the low surface brightness regions. FITS images of $500\times500$ pixels centred on each galaxy were downloaded from the Legacy Survey cutout service at the same interpolated pixel scale as in \citetalias{Walmsley2021}. We then stacked the images by summing the \textit{g}, \textit{r}, and \textit{z} bands, ensuring sensitivity to tidal features regardless of their colour. 

\begin{figure}
    \includegraphics[width=\columnwidth]{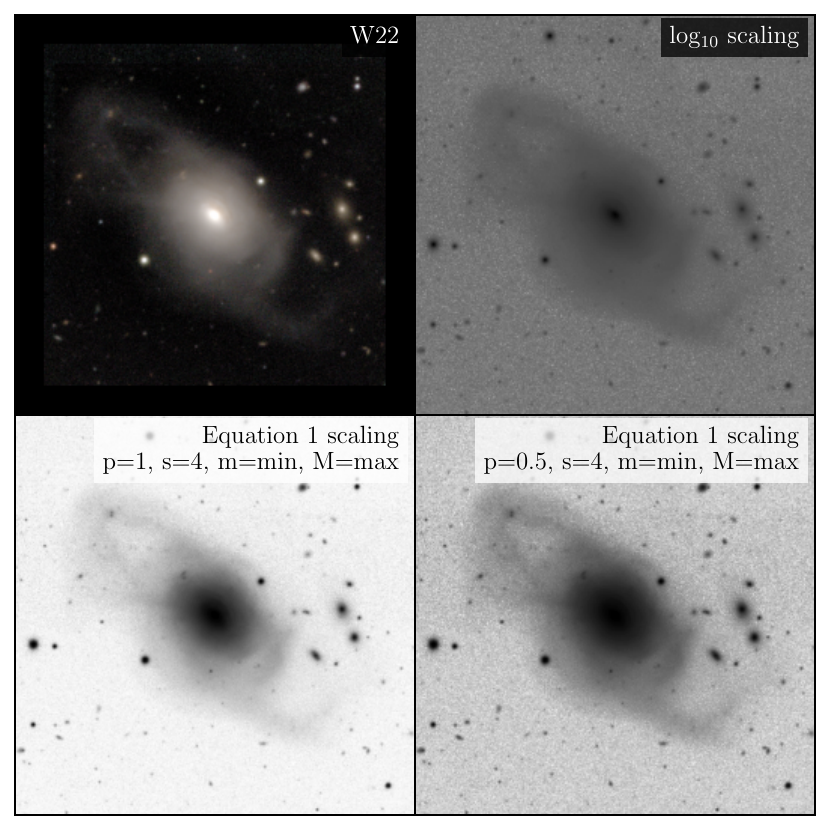}
    \caption{The galaxy J094036.39+033436.9 as an example of the images presented to the inspector during the visual inspection process. The top left shows the image produced by \citetalias{Walmsley2021}, the top right shows the logarithmically scaled image, and the bottom row shows the novel pixel scaling algorithm introduced in this work and described by Equation \ref{eqn:new_arcsinh_scaling} with the parameters $M=\max({\text{pixel values}})$, $m=\min({\text{pixel values}})$, $s=4$, and $p=1$ (left) or $p=0.5$ (right).}
    \label{fig:zooniverse_example}
\end{figure}

We tested various pixel stretch algorithms and selected the three best at enhancing the appearance of the tidal features. These were a logarithmic scaling and two novel arcsinh-based algorithms. The pixel-wise output, $x^{\prime}_{ijk}$, of the novel algorithm depended on the input pixel value $x_{ijk}$ and four other controllable parameters: the maximum $M$, minimum $m$, stretch $s$, and power $p$. The output was then
\begin{subequations}\label{eqn:new_arcsinh_scaling}
\begin{equation}
    x^{\prime}_{ijk} =
    \begin{cases}
        0 & x_{ijk} < m\\
        \frac{\sinh^{-1}\left[ \alpha(x_{ijk};\; s, m, p) \right]}{\sinh^{-1}\left[ \alpha(M;\; s, m, p) \right]} & m \leq x_{ijk} \leq M\\
        1 & x_{ijk} > M
    \end{cases}\\
\end{equation}
where
\begin{equation}
    \alpha(x; s, m, p) = \left(\frac{s|x-m|}{1 + s|x-m|}\right)^p
\end{equation}
\end{subequations} and was inherently constrained to be in the range $\left[0,1\right]$ due to the normalisation. During the visual inspection, we presented the inspector with these three scaled images alongside the original produced by \citetalias{Walmsley2021}. An example of the images presented during the visual inspection is provided in Figure \ref{fig:zooniverse_example}. During training, the network was presented with only the images produced by \citetalias{Walmsley2021} (see Section \ref{sec:CNNs}).

\subsection{Inspection Process}\label{sec:inspection_method}

\begin{figure}
    \includegraphics[width=\columnwidth]{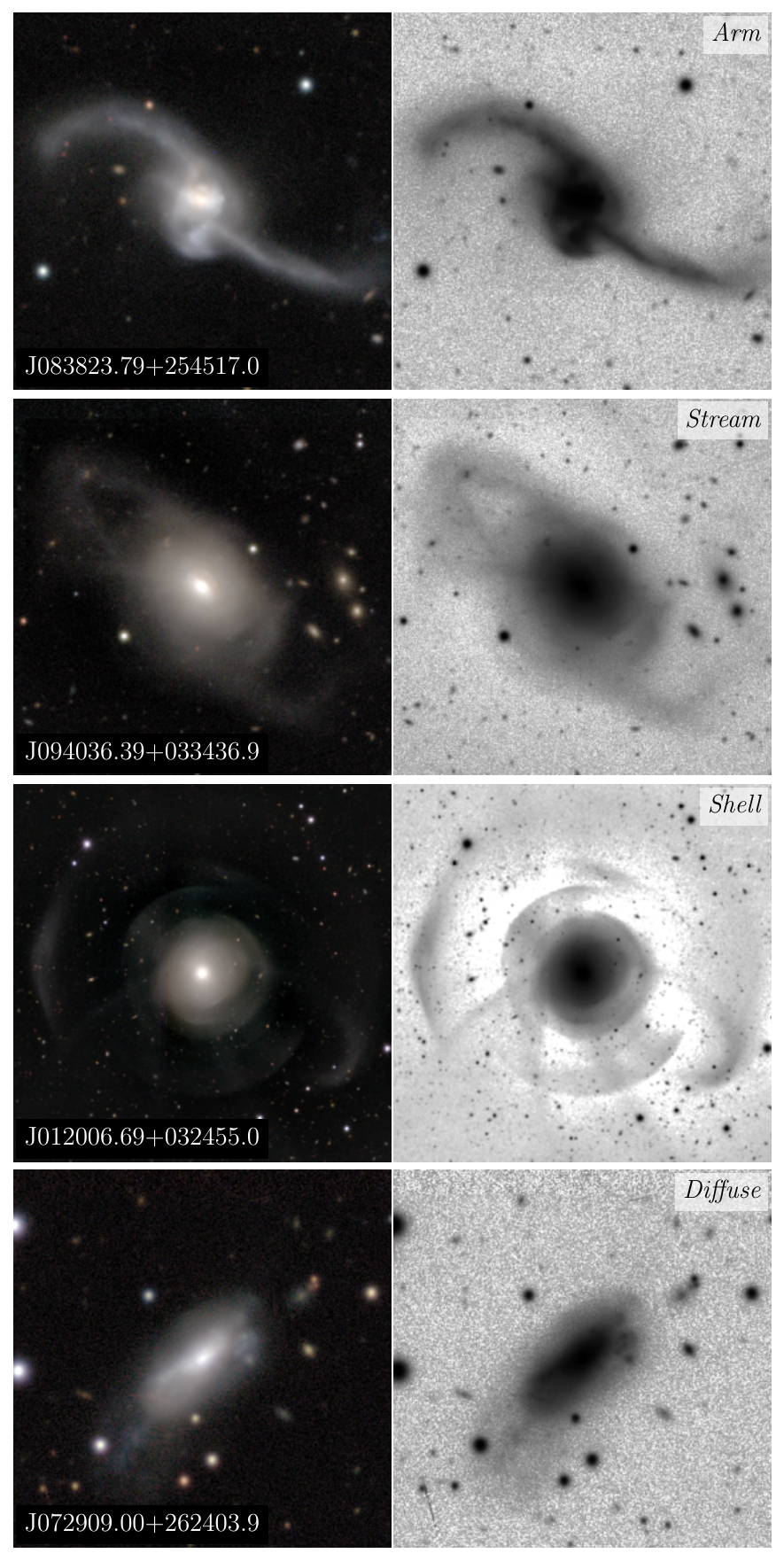}
    \caption{Examples of each of the four categories of tidal features that we visually searched for in the DECaLS sample: \textit{arm}, \textit{stream}, \textit{shell}, and \textit{diffuse}. Each inspector would identify which features were present on the image, selecting all that were visible. Left-hand images were created by \citetalias{Walmsley2021} and were used to train the network described in Section \ref{sec:architecture}, right-hand images show the same galaxy using the Equation \ref{eqn:new_arcsinh_scaling} pixel scaling with the parameters $M=\max({\text{pixel values}})$, $m=\text{med}({\text{background}}) - \text{stdev(background)}$, $s=4$, and $p=0.5$, where the background was estimated using a sigma clipping with a $3\sigma$ clipping limit.}
    \label{fig:example_features}
\end{figure}

We used the Zooniverse.org\footnote{\url{https://www.zooniverse.org/}} platform to make and record classifications for each of the 1,928 galaxies likely to show tidal features. Zooniverse has become a valuable tool for detection and classification studies, most notably including citizen scientists \citep{Lintott2008GalaxySurvey} such as the volunteers in \citetalias{Walmsley2021}. In our case, each of the inspectors (the authors) was presented with the four associated images for each galaxy, as described above and shown in Figure \ref{fig:zooniverse_example}. The inspector was then asked to classify the galaxy into five non-exclusive categories: \textit{arm}, \textit{stream}, \textit{shell}, \textit{diffuse}, or \textit{uncertain}. Where there were multiple features of different kinds, the inspector selected all that applied, and where there were several of the same features, the appropriate category was selected only once. Each of the three inspectors provided separate classifications, which were then combined to produce a label for every galaxy.

The choice of these categories was motivated in part to follow similar works in the literature and to have some connection to the potential astrophysical origin of the feature, thus allowing the classification to be driven towards specific science cases. Figure \ref{fig:example_features} provides an example of each type of tidal feature we chose, and a brief description of their characteristics is provided below:

\textit{Arm}: tidal \textit{arms} form from material that originates in the host galaxy, so the surface brightness of these features is generally higher close to the galaxy's main body and tails off with increasing radial distance. They have colours similar to the outer regions of the host galaxy and should be clearly connected to it. We follow the nomenclature of \citet{Atkinson13} and call these features \textit{arms} but note that they are also referred to as \textit{tails} in the literature \citep[see e.g.][]{Bilek2020CensusSurvey, Martin2020GalaxyLearning, Sola2022CharacterizationImages, Desmons2023GalaxySamples}.

\textit{Stream}: \textit{streams} have similar morphologies to \textit{arm} features; however, they are generally not physically connected to the parent galaxy. They are often brightest near the remaining core of the progenitor, with surface brightness falling off as one moves away from this. As they originate from the disruption of a small satellite galaxy, they are usually far fainter than the parent galaxy. These features could be short, linear, or wrapped around the parent galaxy, depending on the type and inclination of the orbit. Most other literature works include a \textit{stream} class \citep{Kado-Fong2018TidalChannels, Bilek2020CensusSurvey, Martin2020GalaxyLearning, Sola2022CharacterizationImages, Desmons2023GalaxySamples} and we note that this class contains the linear class from \citet{Atkinson13}.

\textit{Shell}: \textit{shells} generally have some symmetry to the shape, such as a shell or fan-like body around the host galaxy and have well-defined, often brighter, edges or caustics. They are likely to originate from nearly radial mergers \citep{2018MNRAS.480.1715P}. We combine the shell and fan classes from \citet{Atkinson13}, and similarly to \textit{streams}, the same literature works include a \textit{shell} class.

\textit{Diffuse}: \textit{diffuse} features lack any well-defined symmetry to the feature or do not fit well into any of the other categories. They have reasonably irregular or asymmetric shapes; however, there can be some ambiguity between genuinely low surface brightness irregular galaxies and \textit{diffuse} debris. The contrast between the galaxy's central region and the debris is a critical diagnostic. When this is small, the systems are likely to be genuinely low surface brightness or irregular galaxies, whereas significant contrast is probably more reflective of debris. The diffuse class is similar to those from \citet{Atkinson13}, \citet{Martin2020GalaxyLearning}, and \citet{Desmons2023GalaxySamples} with the same name. Some of the diffuse features may be genuinely diffuse in nature, while others may simply be very low signal-to-noise examples of some of the previously-mentioned classes.

\textit{Uncertain}: when a galaxy could not be reliably classified into one of the above categories, it was labelled \textit{uncertain}. Most of these galaxies consisted of those too faint or small to reliably say any feature was present or those insufficiently distinct from intrinsically irregular galaxies.

We chose not to include some classes from other works such as bridges, merger remnants, or double nuclei \citep{Martin2020GalaxyLearning, Desmons2023GalaxySamples} as our focus concentrates on post-merger features, and these classes likely represent a mid-merger phase.

\subsection{Classification Demographics}\label{sec:inspection_results}

We combined the inspector classifications for every galaxy by considering where the majority (at least 2 out of 3) agreed that a particular feature was present. For example, if one inspector said \textit{arm} + \textit{stream}, another said \textit{arm}, and the last indicated \textit{stream} + \textit{shell}, the resulting label would be \textit{arm} + \textit{stream}. We follow this process for \textit{uncertain} labels, but where a galaxy was given an \textit{uncertain} designation, it was not given any other label regardless of whether it would have received a label corresponding to a tidal feature. Where the individual classifications did not agree on any label, the galaxy was classified as \textit{none}. Thus, we end up with seventeen possible labels based on the various combinations of features -- specifically, 
\textit{none};
\textit{uncertain};
\textit{arm};
\textit{stream};
\textit{shell};
\textit{diffuse};
\textit{arm} and \textit{stream};
\textit{arm} and \textit{shell};
\textit{arm} and \textit{diffuse};
\textit{stream} and \textit{shell};
\textit{stream} and \textit{diffuse};
\textit{shell} and \textit{diffuse};
\textit{arm}, \textit{stream} and \textit{shell};
\textit{arm}, \textit{stream} and \textit{diffuse};
\textit{arm}, \textit{shell} and \textit{diffuse};
\textit{stream}, \textit{shell} and \textit{diffuse};
and \textit{arm}, \textit{stream}, \textit{shell} and \textit{diffuse}.

\begin{figure*}
   \begin{subfigure}[b]{\columnwidth}
        \includegraphics[width=\columnwidth]{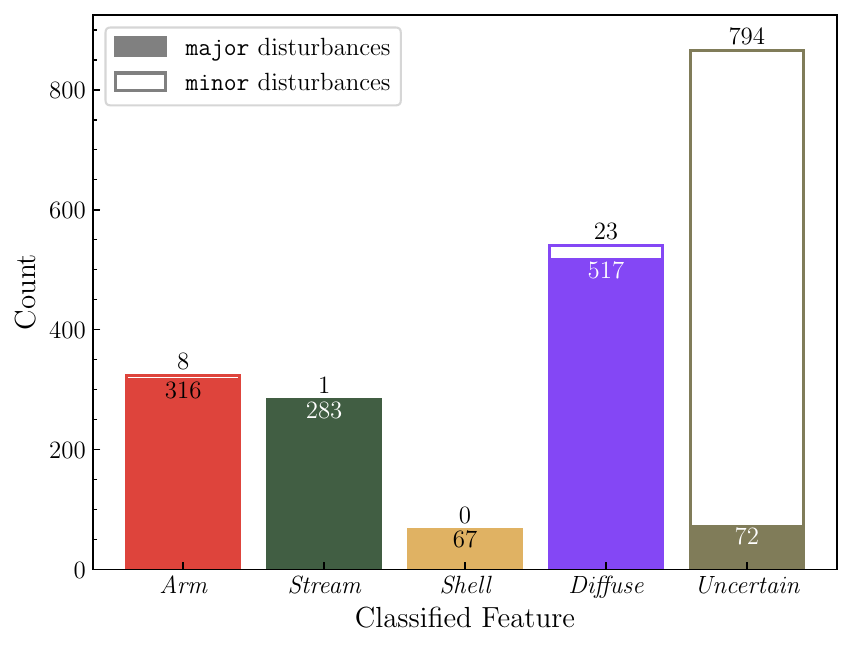}
        \caption{Categories of tidal features.}
        \label{fig:classification_demographics-subfig:features}
    \end{subfigure}
    ~
    \begin{subfigure}[b]{\columnwidth}
        \includegraphics[width=\columnwidth]{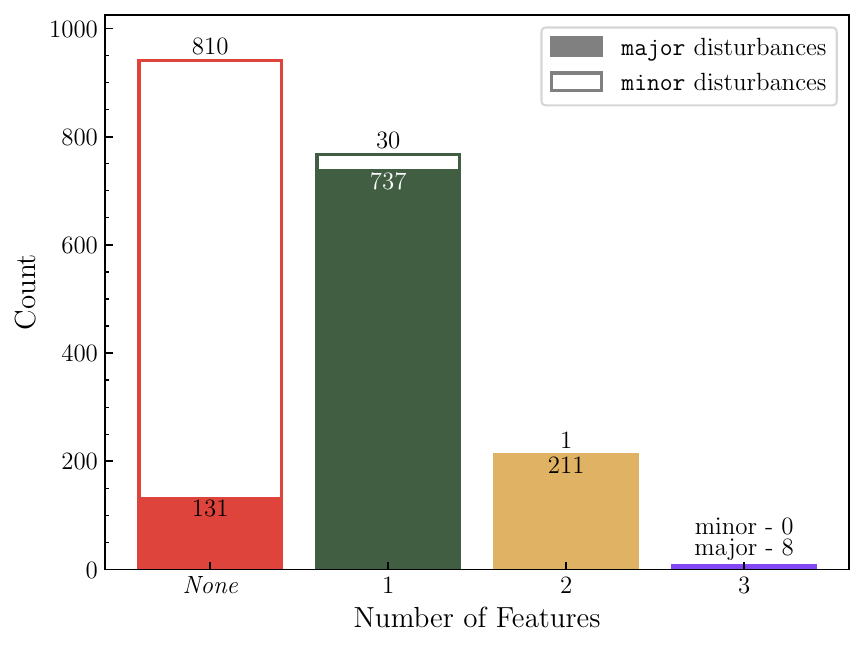}
        \caption{Number of tidal features.}
        \label{fig:classification_demographics-subfig:number}
    \end{subfigure}

    \begin{subfigure}[b]{\columnwidth}
        \includegraphics[width=\columnwidth]{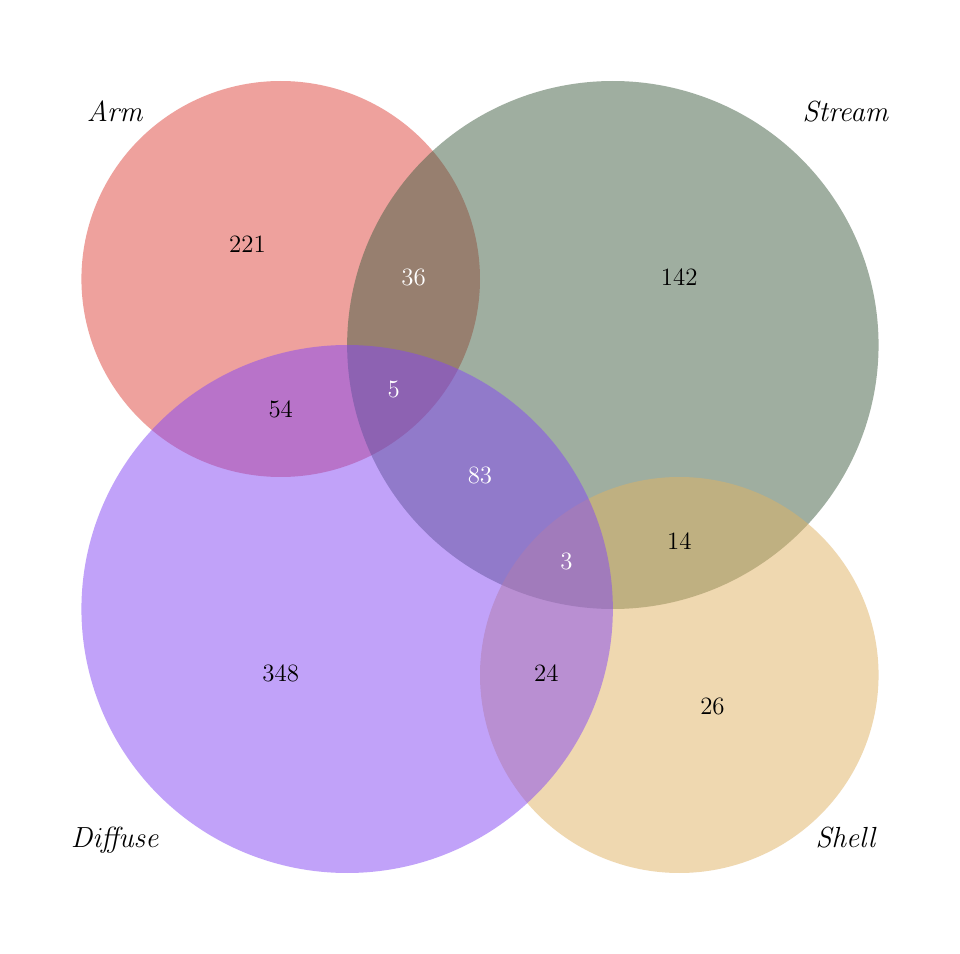}
        \caption{Overlap in classifications of feature categories.}
        \label{fig:classification_demographics-subfig:venn}
    \end{subfigure}
    
    \caption{Demographics of the visual inspection of galaxies for the presence of tidal features. Each galaxy was labelled by considering where two out of three inspectors agreed a given feature was present. (\subref{fig:classification_demographics-subfig:features}) provides the number of galaxies that have \textit{arm}, \textit{stream}, \textit{shell}, \textit{diffuse}, and \textit{uncertain} labels. Galaxies labelled with multiple features are counted multiple times in this Figure. (\subref{fig:classification_demographics-subfig:number}) indicates how many galaxies contained multiple features. Galaxies where no label was agreed on or labelled as \textit{uncertain} are included here as \textit{none}. In both (\subref{fig:classification_demographics-subfig:features}) and (\subref{fig:classification_demographics-subfig:number}), the histograms are split by whether the \citetalias{Walmsley2021} classifier rated the galaxy highly in the merging minor-disturbance fraction or merging major-disturbance fraction columns. Finally, (\subref{fig:classification_demographics-subfig:venn}) shows the overlap between the classifications made. In all three, the numbers shown indicate the number of galaxies in that category.}
    \label{fig:classification_demographics}
\end{figure*}

While the ultimate goal of the visual inspection was to generate labels to train a CNN, it is also of some interest to explore the properties of the classifications. Hence, in Figure \ref{fig:classification_demographics}, we present the demographics of our classifications, split based on whether the prediction of the galaxy was from the \verb|major| (solid) or \verb|minor| (outline with a hollow centre) columns in \citetalias{Walmsley2021}. In both (\subref{fig:classification_demographics-subfig:features}) and (\subref{fig:classification_demographics-subfig:number}), the numbers provided represent the number of galaxies in each category, again divided based on the \citetalias{Walmsley2021} prediction.

Figure \ref{fig:classification_demographics}(\subref{fig:classification_demographics-subfig:features}) provides the number of galaxies with each feature class, noting that this Figure will count galaxies with more than one feature multiple times. Both \textit{stream} and \textit{arm} features appear to be roughly as common as each other (16.8 and 14.7 per cent of the inspected galaxies, respectively), \textit{diffuse} features appear to be the most common (28.0 per cent), and \textit{shells} the least (only 3.5 per cent).

Figure \ref{fig:classification_demographics}(\subref{fig:classification_demographics-subfig:number}) shows the number of features present for each galaxy; \textit{uncertain} galaxies have been included in the \textit{none} category as there was no useable feature present. Most galaxies exhibit a single tidal feature at the surface brightness depth of DECaLS. For some galaxies, all the authors agreed that there was a feature (i.e. not labelled \textit{uncertain}), but no consensus on the type was reached. Hence, more galaxies are listed under \textit{none} (941) than those under \textit{uncertain} (866). Figure \ref{fig:classification_demographics}(\subref{fig:classification_demographics-subfig:venn}) provides the Venn diagram of the classification demographics. We observe that no galaxy contains all four features, and no combination includes both an \textit{arm} and a \textit{shell} feature.

What is of particular note is the \verb|minor| galaxies. As shown in Figure \ref{fig:classification_demographics}, most of the \textit{uncertain} and \textit{none} feature categories are occupied by those indicated as \verb|minor| in \citetalias{Walmsley2021}. Of the 841 galaxies labelled by \citetalias{Walmsley2021} as a minor disturbance (43.6 per cent of the inspected sample), less than 4 per cent had a label that was not \textit{uncertain}. Therefore, we decided to remove all \verb|minor| galaxies entirely and instead create our training sample from the \verb|major| disturbance galaxies only.

We believe this mismatch between the predictions from the \citetalias{Walmsley2021} classifier and our inspection results from a mixture of effects. In most cases, the volunteers rated the galaxy as highly likely to have a \verb|minor| disturbance and the automated classifier followed these labels. However, in some cases, the automated classifier overestimated the fraction of votes for the \verb|minor| disturbance such that it was above our threshold even though the volunteers had rated it unlikely. As a curiosity, we also note that a significant proportion ($\sim$20 per cent) of the \verb|minor| galaxies we remove received fewer than five votes from the volunteers. In addition to removing all of the \verb|minor| galaxies, we excluded those that received an \textit{uncertain} or \textit{none} label. Thus, our final training sample included 956 galaxies with tidal features, about one-half of the 1,928 we initially classified. Nonetheless, this is still one of the largest samples of galaxies with faint tidal debris amassed to date.

\begin{figure*}
    \includegraphics[width=0.9\textwidth]{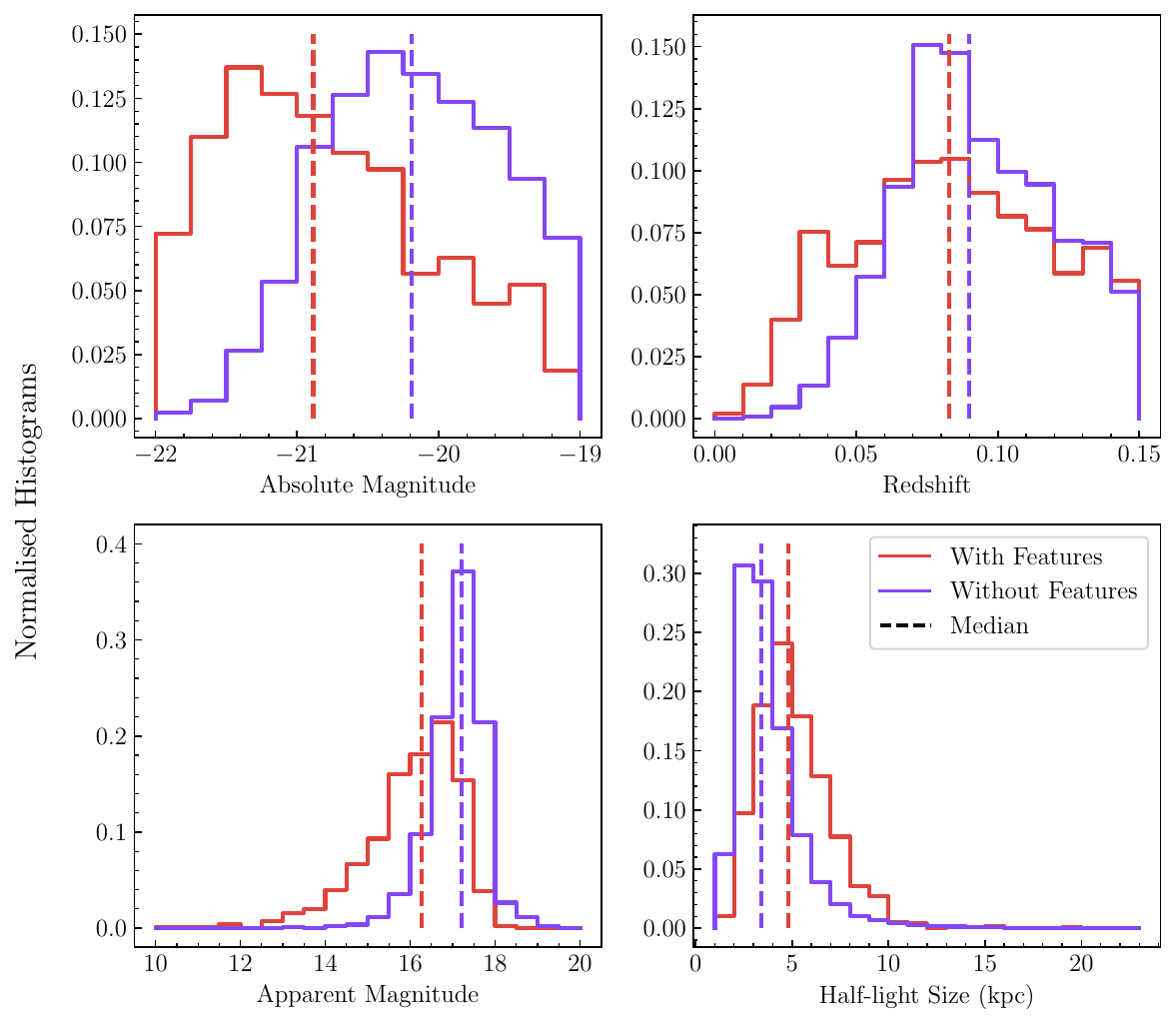}
    \caption{Distributions of galaxy sample properties, including absolute and apparent \textit{r}-band magnitudes, redshift, and half-light size. The distributions are split and normalised based on whether the galaxy had a tidal feature (red) or not (blue). The dashed line provides the median value. All properties are derived from values in the NASA Sloan Atlas (NSA) assuming the \citet{2020A&A...641A...6P} cosmology.}
    \label{fig:sample_properties}
\end{figure*}

Figure \ref{fig:sample_properties} presents the distributions of various properties of the galaxies in the sample split based on whether it had a tidal feature (956, red) or not (9,981, blue). The properties include the absolute and apparent magnitudes in the \textit{r}-band, the redshift, and the size of the galaxies. The sizes of the galaxies were determined by calculating the angular diameter distance from the redshift and combining this with the Petrosian half-light radius. We assumed the \citet{2020A&A...641A...6P} cosmology with $H_0 = 67.66 \;\text{km}\;\text{Mpc}^{-1}\text{s}^{-1}$, $\Omega_{m} = 0.3111$, $\Omega_{\Lambda} = 0.6889$. Each distribution was normalised by the number of galaxies in that category, and the dashed line shows the median value. Within this particular sample, the most noticeable result is that tidally-disturbed galaxies tend to be more intrinsically luminous, although this may be a result of galaxies being removed during our visual inspection if they were too faint or small to reliably classify as having tidal features.

\section{Automated Classification}\label{sec:CNNs}

The visual inspection process took $\sim77$ hours in total for just 1,928 galaxies. This was in part due to the need to identify faint and subtle structures, which may have increased the inspection time compared to other inspection problems (e.g., spiral vs. elliptical). Regardless, this was a significant amount of effort on a relatively small data set, especially compared to the volume of data expected from forthcoming surveys. Therefore, we used our set of visually inspected galaxies to provide a training set to develop an automated process for classifying tidal features with the intention of applying the process to future data later. We chose to use a CNN for our approach.

\subsection{Architecture}\label{sec:architecture}
In essence, a CNN is a model comprising a series of different computations or operations and how these are organised is known as the architecture. The architecture can be considered a series of connected nodes grouped into layers. All the nodes in a layer perform the same operation, each representing where a given operation occurs. The outputs from previous nodes are used as inputs to the next layer of nodes. In a typical CNN used for classification, the layers can be grouped into two parts: a feature extraction part and a classification part. The feature extraction part in a CNN consists of a series of convolutions with various kernels and biases. The goal of the feature extraction part is to generate a latent space (a multi-dimensional embedding) representation of the galaxies, specifically where similar galaxies cluster together within the space \citep[e.g.][]{Alzubaidi2021ReviewDirections}. The network's classification part takes the latent space representation of the image and applies a series of linear combinations with weights and biases to generate a prediction. Overall, the goal is for the network to optimise the parameters of the kernels, weights, and biases such that the output matches the target for the given task as well as possible.

\begin{table}
    \caption{The architecture of the convolutional neural network. The architecture sets out the sequential (top to bottom) order of the layers of different operations, the number of nodes in each layer, the activation functions used, and the number of trainable parameters in that layer. Bracketed numbers show the number of nodes or parameters for the case of the binary network instead of the multi-label classification problem (see Section \ref{sec:binary_results}). See the text for details of the operations performed in each layer.}
    \label{tab:network_architecture}
    \begin{tabular}{l c c c c}
        \hline
        Layer & Nodes & Kernel Size & Activation & Parameters\\
        \hline
        Conv2D & 32 & $3\times3$ & ReLU & 896 \\
        MaxPool2D & -- & $2\times2$ & -- & -- \\
        Conv2D & 48 & $3\times3$ & ReLU & 13,872 \\
        MaxPool2D & -- & $2\times2$ & -- & -- \\
        Conv2D & 64 & $3\times3$ & ReLU & 27,712 \\
        MaxPool2D & -- & $2\times2$ & -- & -- \\
        Flatten & -- & -- & -- & -- \\
        Dense & 64 & -- & ReLU & 3,686,464 \\
        Dropout (rate 0.5) & -- & -- & -- & -- \\
        Dense & 4 (1) & -- & Sigmoid & 260 (65)\\
        \hline
        \multicolumn{2}{l}{Total parameters} & & \multicolumn{2}{r}{3,729,204 (3,729,009)}\\
        \hline
    \end{tabular}
\end{table}

We used a modified version of the network architecture used by \citetalias{Walmsley18} and subsequently adopted by \citetalias{DominguezSanchez2023}. The architecture is provided in Table \ref{tab:network_architecture}, along with the number of trainable parameters associated with that layer, and is described in more detail below. Our aim for this work was not to provide a definitive solution to this problem but rather to demonstrate that it was possible to use CNNs to classify different kinds of tidal features, so we did not further optimise the hyperparameters of the network beyond that of \citetalias{Walmsley18}. Instead, we focused on testing the applicability of the already optimised network to our slightly different task.

The extraction part of the network was composed of three convolutional blocks: a 2D convolution layer with a $3\times3$ kernel, a Rectified Linear Unit (ReLU) activation function, and a $2\times2$ shaped max pooling. The blocks consisted of 32, 48, and 64 nodes, respectively. The convolutional layer works by convolving each channel of the input image (e.g. survey bands) with a kernel. The resulting outputs are then summed over the number of input channels. This process is repeated for all \textit{m} nodes of the convolutional layer. The network often comprises a series of convolutional layers that apply this process to transform the \textit{m} nodes of the previous layer to the \textit{n} of the next \citep[e.g.][]{Goodfellow2016DeepLearning}. Pooling layers reduce the size and training time of the network by replacing values over a specified region with a summary value, such as the maximum or average \citep[e.g.][]{Alzubaidi2021ReviewDirections}. This also induces the network to be invariant to small translations in the input image \citep[e.g.][]{Goodfellow2016DeepLearning}. 

The classification part consisted of two fully connected (or dense) linear layers. Fully-connected layers apply a linear combination of weights and biases to all of the outputs from the previous layer \citep[e.g.][]{Goodfellow2016DeepLearning}. The first layer had 64 nodes and the other was a 4-node output layer -- one node for each category of tidal feature (this was a single node for the binary classification, see Section \ref{sec:binary_results}).

During training, a portion of the data is reserved for validation; this validation data effectively serves as unseen data to test the generalisation of the network. When the metrics used to monitor training performance begin to diverge from those evaluated on the validation data, this can indicate that the network is overfitting the training data. We employed both dropout -- which we applied between both of the fully connected layers -- and early stopping to prevent the network from overfitting to the training data, which would negatively impact the generalisation of the network to unseen (or testing) data. Dropout randomly removes neurons at each step during training with a specified rate \citep{Srivastava2014Dropout:Overfitting}, which we chose to be 50 per cent in line with \citetalias{Walmsley18}. This prevented particular parts of the network from being overly crucial for a given classification and meant the network had to learn many independent features \citep{Alzubaidi2021ReviewDirections}. Early stopping monitors the validation loss metric. It determines whether an improvement has been made based on the overall antecedent minimum value and that determined at the current epoch. After a specified number of epochs -- the patience period -- with no improvement, the network stops training and restores the network parameters that obtained the best value of the validation loss. We chose this patience period to be 40 epochs.

\subsection{Augmentation}\label{sec:augmentation}
We can see from Table \ref{tab:network_architecture} that there were $\sim10^6$ parameters that needed to be constrained through training. This meant we required a large and diverse training set to avoid overfitting the data. Unfortunately, we could not expand our training set beyond the galaxies we had already inspected for tidal features. However, data augmentation allowed us to artificially extend the training set \citep[see, e.g.][]{Goodfellow2016DeepLearning, Shorten2019ALearning, Alzubaidi2021ReviewDirections} without visually classifying further galaxies. The presence or absence of a feature was independent of transformations or augmentations of the input image, such that instances of the same image can be used repeatedly during training. \citetalias{Walmsley18} took this approach when they were training their network on the 305 galaxies inspected by \citet{Atkinson13}. We applied the following augmentations to the data randomly, noting a slight modification from \citetalias{Walmsley18} in the rotation augmentation:
\begin{enumerate}
    \item horizontal or vertical flip or both
    \item rotation in the range $\left[-\frac{\pi}{2}, +\frac{\pi}{2}\right]$
    \item translation up to $\pm$5 per cent along both axes
    \item zoom in or out up to 10 per cent
\end{enumerate}
After these augmentations, we downsized by rebinning the images to $256 \times 256$ pixels. We found that resizing the images improved the network performance on the order of a few per cent and chose $256 \times 256$ as the optimal.

\subsection{Training Sample Construction}\label{sec:fold_construction}

\begin{table*}
    \caption{Composition of each fold for training and testing the network. The galaxies with and without tidal features were divided into five groups. Each fold was constructed to include a testing set comprising one of the five groups and a training set made up of the other four groups. A random sampling process was used to ensure the network was trained on an equal number of galaxies with and without tidal features. Each fold is unique and used to test the network on unseen data, allowing all of the galaxies to be used in testing the network.}
    \label{tab:fold_construction}
    \centering
    \begin{tabular}{l c c c c c c c c c c c c}
        \hline
        \multirow{2}{*}{Fold} & \multicolumn{6}{c}{Training} & \multicolumn{6}{c}{Testing} \\
        & \textit{Arms} & \textit{Streams} & \textit{Shells} & \textit{Diffuse} & \textit{None} & Total & \textit{Arms} & \textit{Streams} & \textit{Shells} & \textit{Diffuse} & \textit{None} & Total \\
        \hline
        Fold 1 & 259 & 227 & 51 & 410 & 765 & 1,530 & 57 & 56 & 16 & 107 & 1,997 & 2,188 \\
        Fold 2 & 254 & 229 & 57 & 414 & 765 & 1,530 & 62 & 54 & 10 & 103 & 1,996 & 2,187 \\
        Fold 3 & 252 & 224 & 51 & 417 & 765 & 1,530 & 64 & 59 & 16 & 100 & 1,996 & 2,187 \\
        Fold 4 & 248 & 228 & 51 & 418 & 765 & 1,530 & 68 & 55 & 16 & 99 & 1,996 & 2,187 \\
        Fold 5 & 251 & 224 & 58 & 409 & 766 & 1,530 & 65 & 59 & 9 & 108 & 1,996 & 2,188 \\
        \hline
        Mean Training Weight & 1.513 & 1.690 & 7.16 & 0.925 & & & & & & & & \\
        $\pm 1\sigma$& 0.024 & 0.017 & 0.47 & 0.009 & & & & & & & & \\
        \hline
    \end{tabular}
\end{table*}

We used our 956 visually inspected galaxies with tidal features and those assumed not to have a tidal feature (9,981, see Table \ref{tab:selection_criteria}) to construct training sets using a five-fold cross-validation scheme (see, e.g. \citealt{Goodfellow2016DeepLearning}; \citetalias{Walmsley18}). In this scheme, we randomly distributed galaxies into five groups; each was unique and roughly equally sized. The random distribution process was performed separately for each type to ensure that each group included both tidal and non-tidal galaxies. To create each fold, we selected one of the five groups to be the testing set and complemented this with a training set. The training set comprised all of the galaxies with tidal features in the other four groups and a random sampling of an equal number of those without, similarly originating from the other groups. Thus, every galaxy in our sample could be used as an unseen testing example to evaluate the performance of our network and our training sets remained balanced between those with and without tidal features. The number of galaxies with each type of feature in each fold is presented in Table \ref{tab:fold_construction}.

\subsection{Multi-Label Results}\label{sec:multi_results}
As with the other works on the subject (e.g. \citetalias{Walmsley18, DominguezSanchez2023, Desmons2023DetectingLearning}), we employ the true positive rate (TPR) or recall, false positive rate (FPR), and area under the curve (AUC) metrics as well as precision, $F_1$, and accuracy to measure the performance of our classifier. Each metric compares some combination of the number of true positives, true negatives, false positives, and false negatives, abbreviated as TP, TN, FP, and FN, respectively. The predictions from the network were separated into positive and negative based on some predictive threshold $P$, and a binary label was assigned for each class based on whether the prediction was greater or lesser than the threshold. These were then compared to labels assigned during the visual inspection to determine whether the predictions were true or false. The FPR 
\begin{equation}
    \text{FPR} = \frac{\text{FP}}{\text{FP} + \text{TN}}
\end{equation}
is a measure of the amount of contamination in the predicted positive samples or the probability a predicted positive example is negative. Conversely, the TPR or recall
\begin{equation}
    \text{TPR} \equiv \text{Recall} = \frac{\text{TP}}{\text{TP} + \text{FN}}
\end{equation}
provides the probability that actual positive examples will be predicted as positive. The AUC score measures the area underneath the curve created by the FPR and TPR values
\begin{equation}
    \text{AUC} = \int^{1}_{0} \text{TPR}\left(\text{FPR}\right) \text{d}\text{FPR}
\end{equation}
for each value of $P$. An AUC of 0 means that the predictions were completely wrong, whereas an AUC of 1 means that the predictions were entirely correct. The precision metric
\begin{equation}
    \text{Precision} = \frac{\text{TP}}{\text{TP} + \text{FP}}
\end{equation} measures the fraction of predicted positives that were correctly predicted. The $F_1$ score
\begin{equation}
    F_1 = 2\frac{\text{Precision} \times \text{Recall}}{\text{Precision} + \text{Recall}} = \frac{2 \text{TP}}{2 \text{TP} + \text{FP} + \text{FN}}
\end{equation} is the harmonic mean of the precision and recall metrics. The accuracy metric
\begin{equation}
    \text{Accuracy} = \frac{\text{TP} + \text{TN}}{\text{TP} + \text{TN} + \text{FP} + \text{FN}}
\end{equation} has its usual meaning as the total fraction of correct predictions.

Each of the folds we created was used to train the network individually. The training set of that fold was input to the training algorithm, with the data split 9:1 for training and validation, respectively. At each step, the images created by \citetalias{Walmsley18} were loaded into the network and randomly augmented. We applied a weighting 
\begin{equation}
    w_{i} = \frac{n_{\text{samples}}}{n_{\text{classes}}n_{i}}
\end{equation} to the training for each class to compensate for the imbalance between the different classes. The weighting was based on the total number of training samples in the set, $n_{\text{samples}}$, divided by the product of the number of classes, $n_{\text{classes}}$, and the number of examples for that class, $n_{i}$. This prioritised training for the higher-weighted classes to compensate for the fewer numbers. Table \ref{tab:fold_construction} provides the mean training weights across the folds. Once the network had been trained, it was used to generate predictions for the testing set of that fold. All memory of the network was erased, and the process was repeated for the next fold. The final result was that each galaxy in our sample had received a prediction from a network where that galaxy had not previously been seen by the network. We then combined these predictions to evaluate the performance of the network. This whole process was repeated for ten trial attempts at training and testing to estimate the error and stochasticity of the performance metrics. Hence, for each galaxy, there were ten unique predictions. We evaluated the performance of the network independently for each of these attempts and then averaged over those performances.

\begin{figure}
    \begin{subfigure}[b]{\columnwidth}
        \includegraphics[width=\columnwidth]{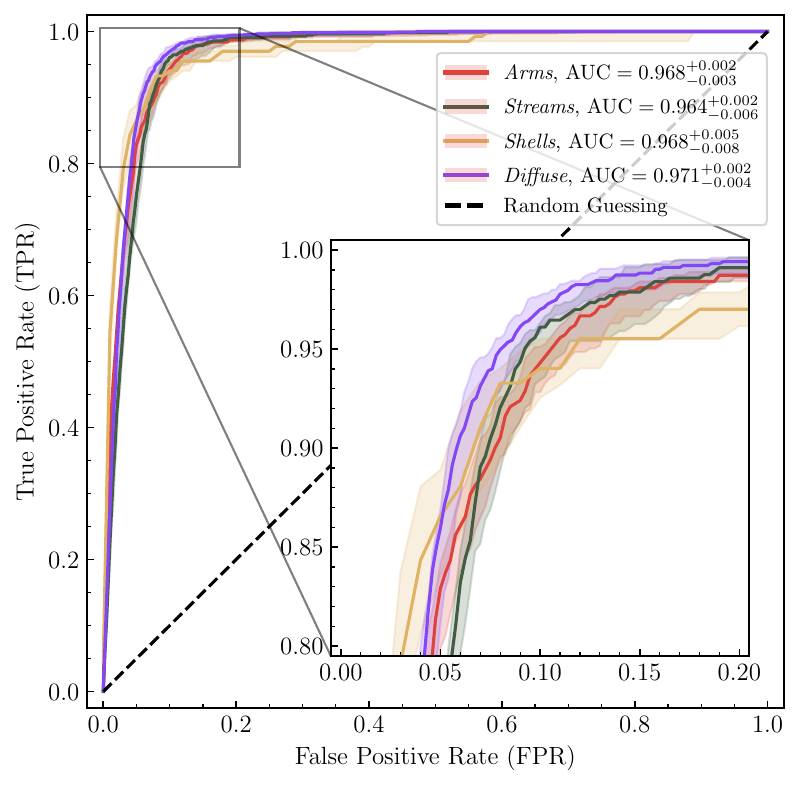}
        \caption{ROC curves for each class}
        \label{fig:multi-label_classification_roc_curve-subfig:classes}
    \end{subfigure}
    
    \begin{subfigure}[b]{\columnwidth}
        \includegraphics[width=\columnwidth]{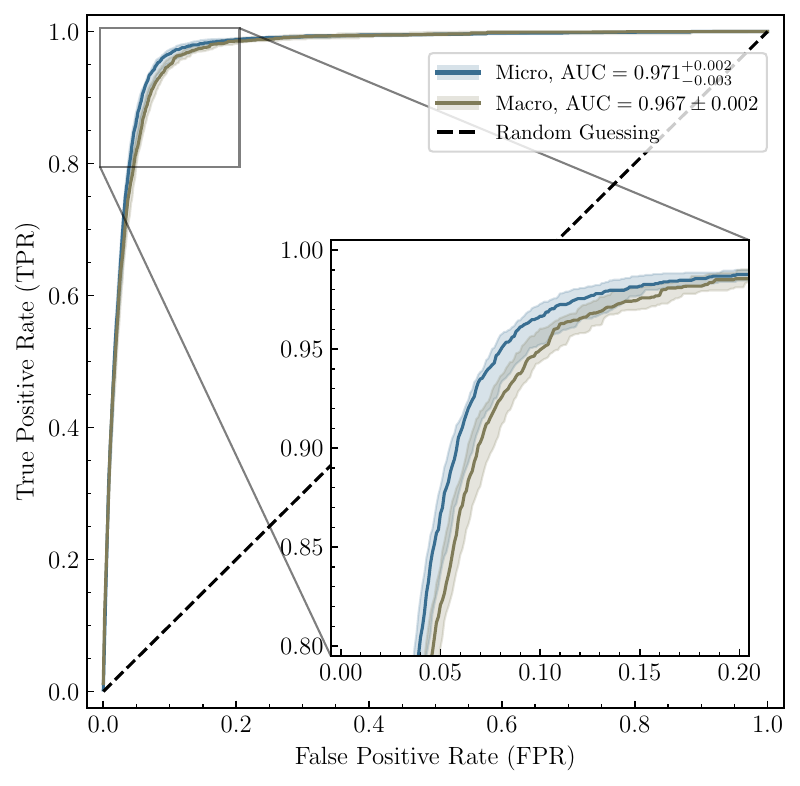}
        \caption{Averaged ROC curves micro (label-wise) and macro (class-wise)}
        \label{fig:multi-label_classification_roc_curve-subfig:avgs}
    \end{subfigure}
    
    \caption{Median Receiver Operating Characteristic (ROC) curves for each of the categories of tidal features (\subref{fig:multi-label_classification_roc_curve-subfig:classes}) and the micro- and macro-averages (\subref{fig:multi-label_classification_roc_curve-subfig:avgs}). The median was calculated by combining the predictions for the testing sets across all five folds and averaging over ten independent trial runs of the network, and the shaded regions indicate the 68.2 per cent confidence interval. The ROC curve indicates how the fraction of correctly identified galaxies (TPR) and the fraction of incorrectly identified galaxies (FPR) change with respect to the predictive threshold -- that being the value above which the output of the classifier would be taken to indicate a prediction the feature was present. The macro-average averages each of the classes' ROC curves, and the micro-average treats each label independently.}
    \label{fig:multi-label_classification_roc_curve}
\end{figure}

As mentioned, each trial network was trained for between 40 and 300 epochs, with the median number of epochs being $196.0_{-48.5}^{+97.5}$. We provide the training and validation loss metric across all five folds and ten attempts in Figure \ref{fig:histories} in Appendix \ref{app:metrics} and verify that no overfitting occurs. 

Figure \ref{fig:multi-label_classification_roc_curve} shows median receiver operating characteristic (ROC) curves evaluated by combining the testing set predictions and labels from each fold and averaging over ten separate trial runs. As the testing sets had roughly one galaxy with a tidal feature for every 10 without, reflective of the true population at this surface brightness depth, the ROC curves were determined from an imbalanced set. The shaded regions in the Figure provide the 68.2 per cent confidence intervals.

The ROC curve shows the relationship between the FPR ($x$-axis) and TPR ($y$-axis) for different predictive thresholds, where lower thresholds result in more completeness but also more contamination. Each class was treated independently because they were non-exclusive, as a galaxy could have more than one feature of different classes. Thus, the ROC curves and other metrics were determined by considering the predictions and labels for each class in a binary fashion (e.g. by comparing only the yes/no label for \textit{arms} to the predicted value for \textit{arms}). The exception is for micro-averaged values, which compare every prediction to its label, regardless of the class. The macro-average instead averages over the metrics evaluated on each class, where each class was weighted the same. Figure \ref{fig:multi-label_classification_roc_curve}(\subref{fig:multi-label_classification_roc_curve-subfig:classes}) shows the ROC curve for each class -- where each class is treated independently -- and (\subref{fig:multi-label_classification_roc_curve-subfig:avgs}) provides the micro- and macro-averages. It is very encouraging to see that the ROC curves are well separated from the random guessing line, corresponding to the TPR and FPR being equal across all thresholds.

\begin{table*}
    \bgroup
    \def\arraystretch{1.5}
    \caption{Median performance metrics and 68.2 per cent confidence intervals. The median was determined by combining the predictions for each fold and averaging over ten trial runs of the network. The metrics are shown for each category of tidal feature, as well as the micro- (equal galaxy weighted average) and macro-averages (equal class weighted average) and the binary classification (see Section \ref{sec:binary_results}). The TPR was evaluated at a fiducial FPR of 0.2. The accuracy, precision, recall, and $F_1$ score metrics were evaluated at an optimal threshold, $P_{\text{opt}}$, which was selected to maximise the $F_1$ score. The $P_{\text{opt}}$ value for the binary classification was taken to be 0.5 rather than the maximising value.}
    \centering
    \label{tab:tpr_values}
    \begin{tabular}{l c c c c c c}
        \hline
        Feature & $\text{TPR}(\text{FPR} = 0.2)$ & $P_{\text{opt}}$ & Accuracy & Precision & Recall & $F_1$ \\
        \hline
        \textit{Arms} & $0.987\pm0.003$ & 0.392 & $0.964^{+0.003}_{-0.005}$ & $0.421^{+0.024}_{-0.034}$ & $0.617^{+0.025}_{-0.045}$ & $0.499^{+0.007}_{-0.020}$
        \\
        \textit{Streams} & $0.991\pm0.005$ & 0.288 & $0.951^{+0.004}_{-0.006}$ & $0.306^{+0.009}_{-0.025}$ & $0.700^{+0.046}_{-0.051}$ & $0.419\pm0.015$
        \\
        \textit{Shells} & $0.970\pm0.008$ & 0.523 & $0.992\pm0.001$ & $0.345^{+0.070}_{-0.025}$ & $0.336^{+0.078}_{-0.076}$ & $0.360^{+0.018}_{-0.070}$
        \\
        \textit{Diffuse} & $0.994^{+0.002}_{-0.006}$ & 0.315 & $0.950^{+0.002}_{-0.006}$ & $0.482^{+0.010}_{-0.033}$ & $0.811^{+0.039}_{-0.013}$ & $0.607^{+0.008}_{-0.032}$
        \\
        Micro & $0.988^{+0.002}_{-0.003}$ & 0.314 & $0.900^{+0.005}_{-0.008}$ & $0.387^{+0.012}_{-0.028}$ & $0.734^{+0.039}_{-0.020}$ & $0.510^{+0.005}_{-0.029}$
        \\
        Macro & $0.986\pm0.004$ & 0.315 & $0.961^{+0.002}_{-0.004}$ & $0.349^{+0.011}_{-0.029}$ & $0.687^{+0.047}_{-0.021}$ & $0.463^{+0.007}_{-0.032}$
        \\
        Binary & $0.9984^{+0.0005}_{-0.0011}$ & 0.5 & $0.981^{+0.001}_{-0.003}$ & $0.834^{+0.014}_{-0.026}$ & $0.974^{+0.008}_{-0.004}$ & $0.900^{+0.073}_{-0.015}$
        \\
        \hline
    \end{tabular}
    \egroup
\end{table*}

Although the performance is generally excellent across all classes, some small variations can be seen. In particular, \textit{Shell} features were recovered at a lower rate than other classes, with the network obtaining a median TPR of $0.970\pm0.008$ for a fiducial contamination of 20 per cent (i.e. at an FPR of 0.2). In contrast, \textit{diffuse} features were recovered well with a median TPR of $0.994^{+0.002}_{-0.006}$ for the same degree of contamination. Table \ref{tab:tpr_values} provides the median TPR values for each class and the micro- and macro-averages at the reference 20 per cent contamination.

\section{Discussion}\label{sec:discussion}

Our results are the first demonstration of using deep learning to classify different types of tidal features. This presents an excellent opportunity for forthcoming imaging surveys such as LSST \citep{Ivezic2019LSST:Products} and Euclid \citep{Laureijs2011Euclid:Universe} to investigate the occurrences of the different species of tidal feature and thus, hopefully, shed light on the formation mechanisms and provide substantial data for population studies. As our primary focus for this work was to demonstrate the application of CNNs to classifying different categories of tidal features, we leave investigations of the properties and populations of these features to subsequent works. Although we observed a very good performance for our network from the ROC curves, we do note that there are issues and some improvements needed which are discussed below. 

\begin{figure}
    \centering
    \includegraphics[width=\columnwidth]{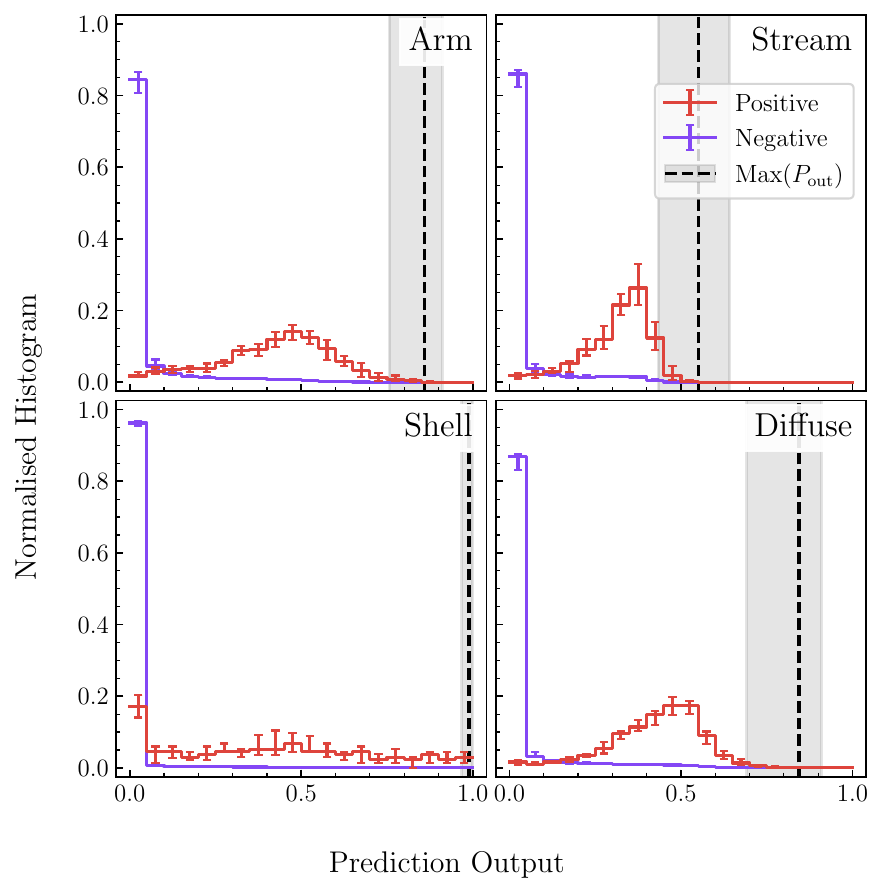}
    \caption{Histograms of the output predictions combined across all five folds. The histogram is split based on whether the truth label for that class is positive (red) or negative (blue). Each value shows the median fraction of positive and negative galaxies with predictions in that bin, and the error bar shows $\pm1\sigma$. The black dashed line and shaded region shows the median $\pm1\sigma$ maximum output prediction across all ten trials.}
    \label{fig:output_distributions}
\end{figure}

One of the first issues we noted was that the output predictions did not always reach a maximum of 1. If the network's output is considered as a probability that a feature is present, then the network is not confident that a feature is there. We expect the output values to match the labels as best as possible, so they should be 1 where the label is 1 (i.e., the binary indicator for that feature being present). However, this aspect of the performance is not captured in the ROC curves seen in Figure \ref{fig:multi-label_classification_roc_curve}. The ROC curves were determined by selecting values of the predictive threshold between $0$ and the maximum output such that the ROC curve starts at $(0,0)$ and ends at $(1,1)$. To gain further insight into the performance of the network, we provide the distribution of output values in Figure \ref{fig:output_distributions} split between those with each type of feature (blue) and those without (red). Also shown in Figure \ref{fig:output_distributions} is the median and 68.2 per cent confidence interval of the maximum output prediction from the network. As mentioned, this does not always reach the expected value of 1 and varies depending on the class, with \textit{streams} having the lowest maximum output of $0.552^{+0.045}_{-0.067}$ and \textit{shells} the highest at $0.989^{+0.006}_{-0.019}$.

\subsection{Variations in class performance}\label{sec:class_performance}

The performance of our network varied across the features being detected; additionally, this variation was different when considering different metrics. For example, looking at the AUC scores in Figure \ref{fig:multi-label_classification_roc_curve}(\subref{fig:multi-label_classification_roc_curve-subfig:classes}), it would appear that the \textit{stream} class is marginally poorer in terms of performance compared to the others. However, from the TPR rates at a fiducial FPR of 0.2 in Table \ref{tab:tpr_values} it would appear that the \textit{shell} class had a slightly poorer performance than the others. In both AUC and $\text{TPR}(\text{FPR}=0.2)$ terms, the \textit{diffuse} class appears to be the best-performing, however the significance is $<3\sigma$.

\begin{figure*}
    \begin{subfigure}[b]{\columnwidth}
        \includegraphics[width=0.95\columnwidth]{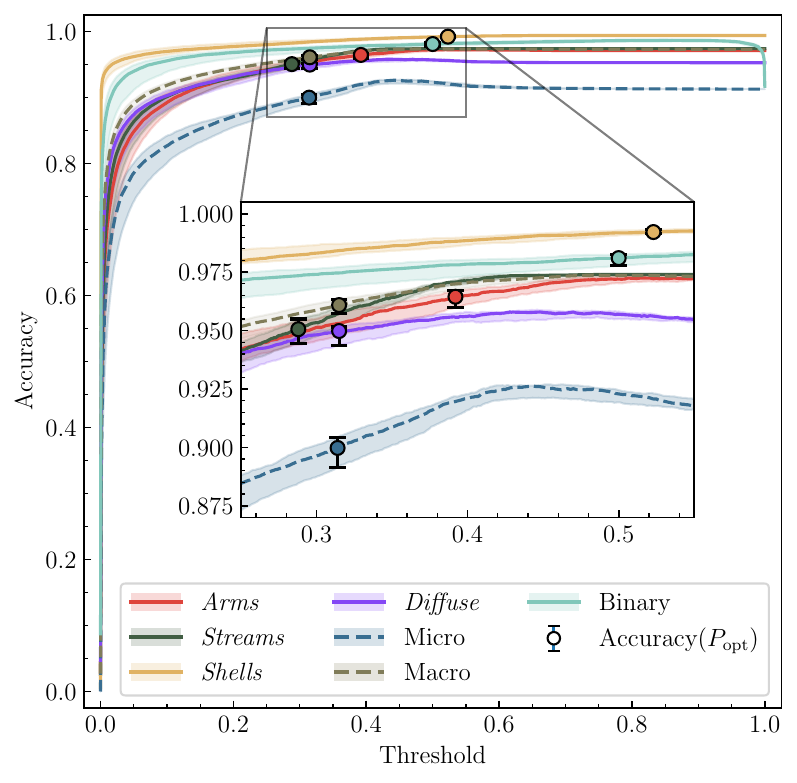}
        \caption{Accuracy}
        \label{fig:classification_performances-subfig:accuracy}
    \end{subfigure}
    ~
    \begin{subfigure}[b]{\columnwidth}
        \includegraphics[width=0.95\columnwidth]{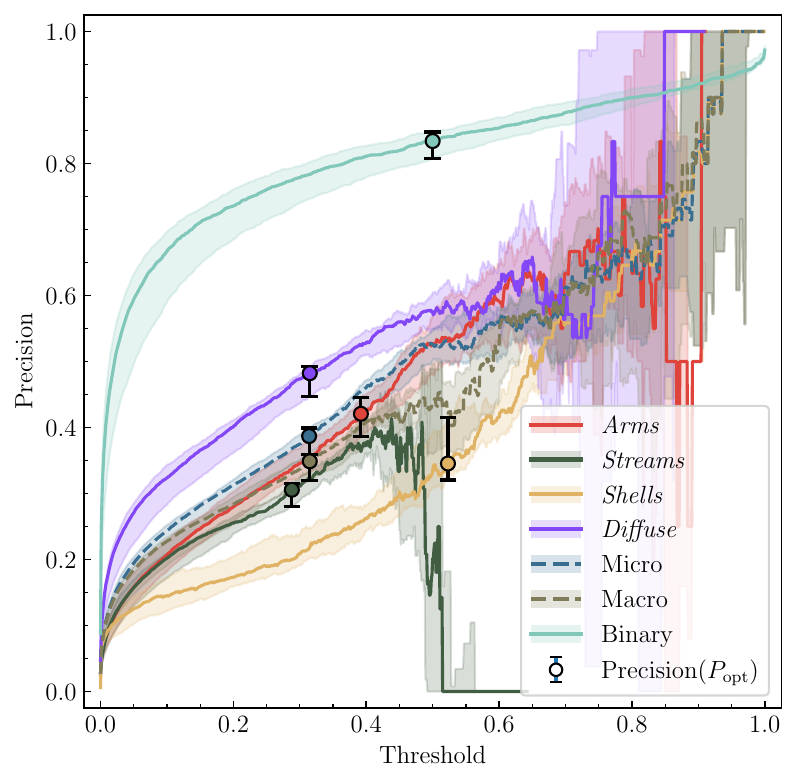}
        \caption{Precision}
        \label{fig:classification_performances-subfig:precision}
    \end{subfigure}

    \begin{subfigure}[b]{\columnwidth}
        \includegraphics[width=0.95\columnwidth]{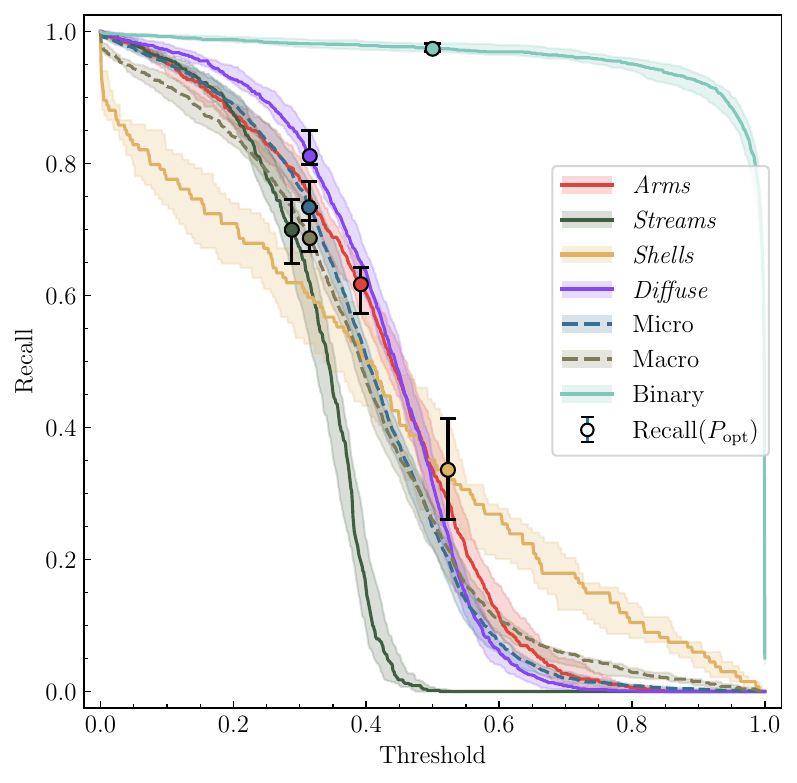}
        \caption{Recall}
        \label{fig:classification_performances-subfig:recall}
    \end{subfigure}
    ~
    \begin{subfigure}[b]{\columnwidth}
        \includegraphics[width=0.95\columnwidth]{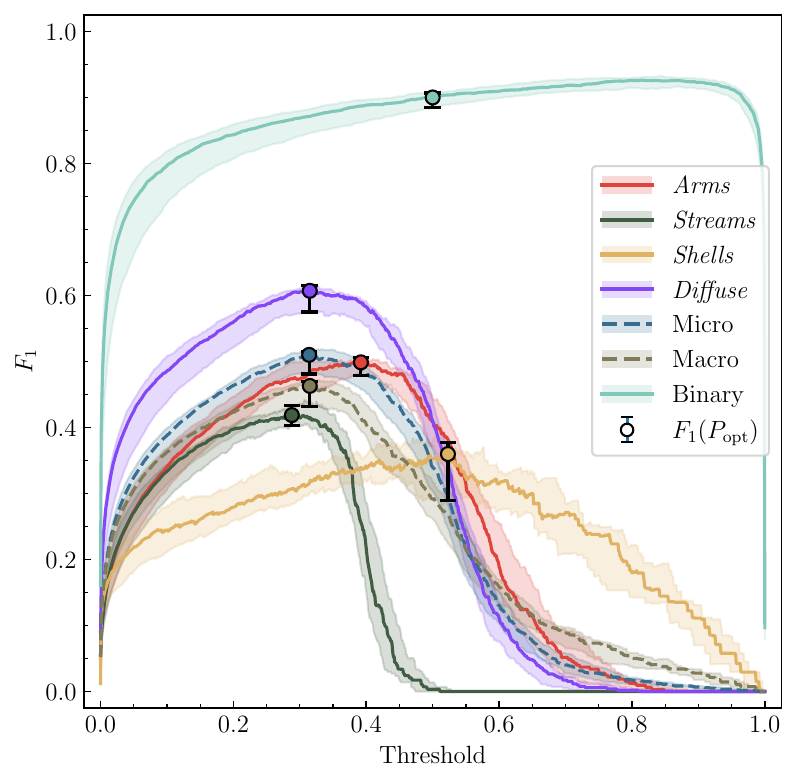}
        \caption{$F_1$ score}
        \label{fig:classification_performances-subfig:f1score}
    \end{subfigure}
    
    \caption{
    The accuracy (\subref{fig:classification_performances-subfig:accuracy}), precision (\subref{fig:classification_performances-subfig:precision}), recall (\subref{fig:classification_performances-subfig:recall}), and $F_1$ (\subref{fig:classification_performances-subfig:f1score}) metrics across various predictive thresholds. The metrics were determined by combining the predictions for every galaxy across all five folds and splitting these into positive and negative based on the predictive threshold. These were then compared to the ground truth labels assigned to the galaxy for each of the given classes. The figure shows the median of ten attempts with the shaded area providing the 68.2 per cent confidence interval. Filled circles provide the value of each metric at the optimal threshold, $P_{\text{opt}}$, as provided in Table \ref{tab:tpr_values}.
    }
    \label{fig:classification_performances}
\end{figure*}

In Figure \ref{fig:classification_performances} we present the accuracy (\subref{fig:classification_performances-subfig:accuracy}), precision (\subref{fig:classification_performances-subfig:precision}), recall (\subref{fig:classification_performances-subfig:recall}), and $F_1$ (\subref{fig:classification_performances-subfig:f1score}) metrics for each class evaluated at a range of predictive thresholds in the range $\left[0,1\right]$ and averaged across the ten trials, along with the 68.2 per cent confidence intervals. Shown too are the micro- and macro-averages of these metrics. We note that results with a threshold beyond the maximum output predictions (shown in Figure \ref{fig:output_distributions}) are meaningless, and hence the metrics are truncated at the maximum output for their respective class. We estimated an optimal threshold, $P_{\text{opt}}$, that maximised the $F_1$ score and show each of the metrics evaluated at the respective $P_{\text{opt}}$ as points on the Figure with the associated error. These values are also provided in Table \ref{tab:tpr_values}.

We again observed that the \textit{diffuse} class appeared to be the best-performing in terms of precision, recall, and $F_1$. Interestingly, on average across all three of these metrics, the network performance correlates with the number of galaxies in that class. Indeed, the \textit{diffuse} class is the best-performing and had the most galaxies (517), and similarly \textit{shells} was the worst with the least (67).

Conversely, the performance appeared to be roughly anti-correlated with the sample size when considering instead the accuracy metric. This likely resulted from the high fraction of galaxies without tidal features in the testing sample, most of which were then classified correctly. Although the network was trained on a $1:1$ ratio, galaxies without tidal features outnumbered those with tidal features at a rate of around $\sim10:1$ in the testing set. We observed that $71.5^{+6.0}_{-6.3}$, $82.8^{+4.2}_{-5.5}$, $66.7^{+8.8}_{-11.6}$, and $60.5^{+3.3}_{-7.8}$ per cent of the false positives for the \textit{arm}, \textit{stream}, \textit{shell}, and \textit{diffuse} classes, respectively, had some other type of tidal feature. Together with the high accuracy scores, this indicates that the network was confident at a binary classification (i.e. distinguishing where some kind of tidal feature is present or not; this is explored further in Section \ref{sec:binary_results}) but sometimes struggled to determine the type of feature, leading to the low precision scores (at low thresholds, see Figure \ref{fig:classification_performances}(\subref{fig:classification_performances-subfig:precision})).

We investigated if the performance of the network had any dependence on the angular size or apparent brightness of the galaxies. We observed that the distribution of false positives was slightly skewed towards larger angular sizes than the distribution of galaxies without tidal features. This small offset could have various origins, such as a bias in the network or in the visual inspection process that created the labels. On the other hand, the effect could reflect a genuine feature, with tidally-disturbed galaxies in our sample being somewhat larger than the overall population. Further exploration of this offset is beyond the scope of the current work.

Furthermore, we consider that the visual inspection labels may have introduced some noise. We often found it difficult during the visual inspection process to separate instances of the two classes as the defining features of the classes are very similar, often with the inspector having to make some estimation of the likely origin of the feature. In particular, inspectors frequently disagreed on whether a feature was a \textit{stream} or \textit{arm}. This suggests that these classes could be combined or the visual inspection process improved to ensure a clearer distinction between the classes, such as more explicit instructions on the differences. We suspect that the network may have been susceptible to this uncertainty in the actual label. We found $25.3^{+3.9}_{-2.5}$ per cent of the false \textit{arm} positives had \textit{streams}, and similarly $27.0^{+1.3}_{-2.1}$ per cent of the \textit{stream} false positives had \textit{arms}. Although these are significant, to put that in context, over fifty per cent of the false positives for the \textit{arm} and \textit{stream} classes included \textit{diffuse} features. We investigated the impact of improving the classification scheme to reduce label noise, and although limited in scope, there was no significant improvement made given the time cost for an entire re-classification of the whole sample (see Appendix \ref{app:improve_classifications}). Finally, we used our network to generate predictions for the \textit{uncertain} galaxies we excluded from the training sets (see Section \ref{sec:inspection_results}). As expected, these galaxies generated a broad distribution of predictions both predicted to have and not have tidal features, reinforcing our strategy of excluding them from training.

\subsection{Comparison to the Literature}
In this Section, we compare our results to those from the literature. First, we default back to comparing the overall binary classification of our sample to match the work already undertaken on other samples. We then compare our multi-label results to these binary classifications of the literature. Finally, we identify whether streams in the DECaLS area identified by \citet{Martinez-Delgado2023HiddenSurvey} were identified by our network.

\subsubsection{Binary Results}\label{sec:binary_results}
To begin the comparison of our work to that of \citetalias{Walmsley18}, \citetalias{DominguezSanchez2023}, and \citetalias{Desmons2023DetectingLearning}, we performed a binary classification. That is, where we considered only if a tidal feature was present on the image regardless of what that tidal feature was. We constructed a binary training set by taking the five folds created earlier (see Section \ref{sec:fold_construction}) and reassigning a positive label to galaxies with any kind of tidal feature. Similarly, those galaxies without tidal features were given a negative label. As with the multi-label scenario, the training sets had a roughly equal number of galaxies with and without tidal features, whereas the testing sets reflected a more realistic number of each.

As before, we preprocess the data by applying random augmentations and resizing the images to be $256 \times 256$. We trained each fold with ten different trial runs of the network for a maximum of 300 and a minimum of 40 epochs, with the median being $226.5^{+44.1}_{-38.1}$. The network was trained using the training portion of the fold, and the network's performance was evaluated on the unseen testing set.

\begin{figure}
    \includegraphics[width=\columnwidth]{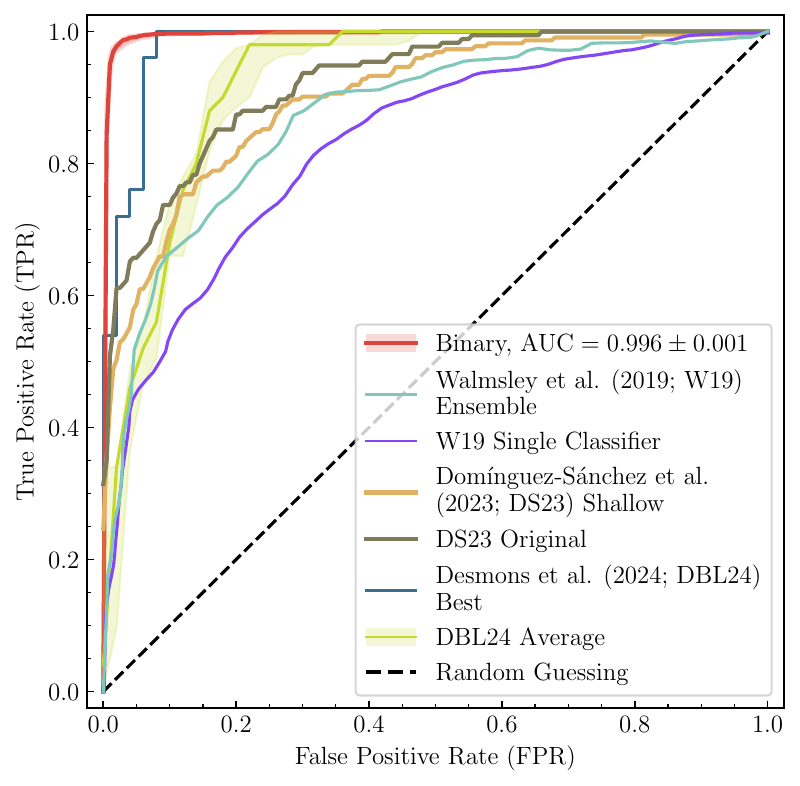}
    \caption{Median ROC curve for the binary classification with comparative literature results. The binary classification considers only positive and negative incidences of any tidal feature, unlike the multi-label classification, which aims to identify the type of tidal feature. Shaded regions indicate the 68.2 per cent confidence interval; again, the median was determined over ten network trials. We include both the single and best-performing ensemble classifiers from \citetalias{Walmsley18}, both shallow and original samples from \citetalias{DominguezSanchez2023}, and the best-performing classifier and our determination of the median of other classifiers from \citetalias{Desmons2023DetectingLearning}.}
    \label{fig:binary_classification_roc_curve}
\end{figure}

Figure \ref{fig:binary_classification_roc_curve} shows the median ROC curve for the binary classification; again, the shaded area shows the 68.2 per cent confidence interval. It is evident from the Figure that our classifier has an outstanding performance, reaching an AUC score of $99.6\pm0.1$ per cent. ROC curves from \citetalias{Walmsley18}, \citetalias{DominguezSanchez2023}, and \citetalias{Desmons2023DetectingLearning} are also shown in the Figure. We also include the accuracy, precision, recall, and $F_1$ metrics in Figure \ref{fig:classification_performances} and Table \ref{tab:tpr_values}. The value of $P_{\text{opt}}$ for the binary classification was taken to be 0.5 rather than that which maximises the $F_1$ score. Considering the performances in these metrics, the network is clearly excellent at distinguishing galaxies with and without tidal features.

However, we note that our binary sample is likely to be a highly idealised scenario. The sample has already been split according to a binary classification schema during the sample selection process, where we selected only galaxies where the \citetalias{Walmsley2021} classifier indicated that there may be a potential tidal feature. Thus, we expect images where the network may struggle to predict whether a tidal feature is present or not to have been excluded from the binary classification set and that the performance on unbiased data could potentially be worse. Our multi-label set represents a new classification schema, so we expect this problem to be a lesser issue as those images have not been selected as good or bad at identifying specific categories of tidal features.

\subsubsection{Multi-Label Comparison}
\begin{figure}
    \includegraphics[width=\columnwidth]{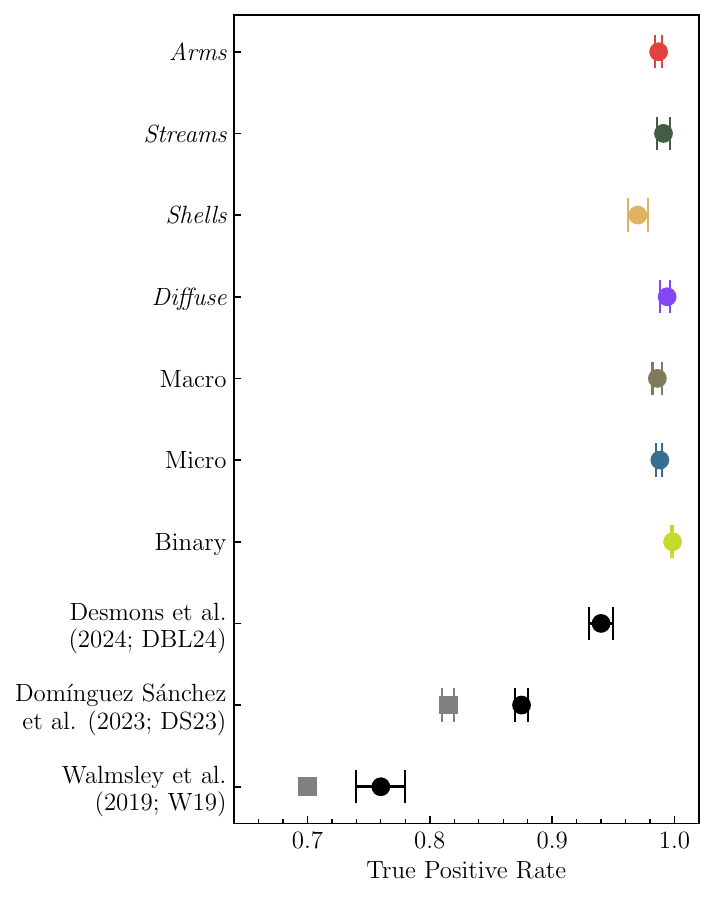}
    \caption{Median TPR at a specified FPR of 0.2 for each class, micro- and macro-average, and a binary classification identifying any tidal feature. The median was calculated by combining the outputs of the five folds and then averaging over ten trial runs of the network, and the error bar represents the 68.2 per cent confidence interval. Additionally shown are the corresponding values from the binary classifiers of \citetalias{Desmons2023DetectingLearning}, \citetalias{DominguezSanchez2023} for both the network trained with (grey square) and without (black circle) shallower data, and \citetalias{Walmsley18} for both a single CNN (grey square) and the average of two ensembles of CNNs (black circle). See the text for details.}
    \label{fig:tpr_comparison}
\end{figure}

Figure \ref{fig:tpr_comparison} presents the median TPR values for each category of tidal feature, the macro- and micro-averages, and the binary classification (from Section \ref{sec:binary_results}) for an FPR of 0.2, as well as the equivalent for \citetalias{Walmsley18}, \citetalias{DominguezSanchez2023}, and \citetalias{Desmons2023DetectingLearning}. We note that the \citetalias{Walmsley18} values were reported at an FPR of 0.22 and the \citetalias{DominguezSanchez2023} values were estimated by reading the appropriate figure. We see in Figure \ref{fig:tpr_comparison} that, in terms of the TPR at an FPR of 0.2, the performance of our networks at identifying the tidal feature classes was comparable to the network trained by \citetalias{Desmons2023DetectingLearning}.

It is important to recall, however, that the detection of tidal features depends heavily on the limiting surface brightness magnitude of the survey \citep[e.g.][]{Johnston2008, Casanova21}. Thus, we expect that the tidal features are more clearly visible and delineated in deeper data, and as such, comparing network performances on varying depth data should take this into account. Indeed, the expectation that networks trained using deeper data will have better performance is consistent with the result of our comparison when considering the limiting surface brightness ($\mu_r \sim28.8$ this work; $\mu_r \sim 27.1$ \citetalias{Walmsley18}; $\mu_r = 26-35$ \citetalias{DominguezSanchez2023}; $\mu_r \sim 29.82$ \citetalias{Desmons2023DetectingLearning}; in mag arcsec$^{-2}$).

\subsubsection{Stellar Stream Legacy Survey}

\citet{Martinez-Delgado2023HiddenSurvey} introduced the Stellar Stream Legacy Survey to identify tidal features around galaxies in the local universe. They used data from the DESI Legacy Imaging Survey DR8, which includes DECaLS, as well as the Beijing-Arizona Sky Survey and Mayall z-band Legacy Survey \citep[BASS and MzLS, respectively;][]{Dey2019OverviewSurveys}. They re-reduced the raw Legacy Surveys data using a bespoke sky subtraction algorithm \citep[see][for details]{Martinez-Delgado2023HiddenSurvey} and then visually inspected a sample of 689 galaxies. From this, they selected 24 galaxies with tidal disturbances -- explicitly chosen to cover a range of morphologies, surface brightnesses, and distances -- and used \textsc{noisechisel} \citep{Akhlaghi2015} to confirm the detections. It is important to note that \citet{Martinez-Delgado2023HiddenSurvey} focus on identifying any type of tidal disturbance that originates from the accretion of a low mass dwarf galaxy and that they do not attempt to separate their sample into different tidal feature classes.

We investigated the overlap between the 24 galaxies they identified and our sample. Of the 24 galaxies in their sample, we find only 18 of these were imaged in DECaLS DR5. Furthermore, 4 of these galaxies were not imaged with SDSS and, as such, were excluded from our sample due to the \citetalias{Walmsley2021} selection criteria. None of the 14 remaining galaxies identified by \citet{Martinez-Delgado2023HiddenSurvey} were included in the galaxies we visually inspected because they did not have sufficiently high tidal disturbance scores from \citetalias{Walmsley2021}. As a reminder, our analysis focused on systems that had a \verb|major| disturbance prediction from the \textit{GZD} automated classifier that was greater than 0.4 whereas these 14 systems had predictions somewhere between 0.08 and 0.4. The reason why they did not score higher is beyond the scope of this work but could plausibly be due to genuine uncertainty in the volunteer classification, the size of the thumbnails presented to volunteers (e.g., too small to include the tidal feature, which in several cases lies at a large distance from the main galaxy) or inaccuracies in the classifier.

Although none of these 14 galaxies were inspected by us, they had high enough prediction scores to be excluded from our assumed undisturbed sample. Thus, we could at least use our network to generate predictions for each of these galaxies to discover if our classifier agrees that they are tidally disturbed. Encouragingly, all but two of the galaxies were identified by our network to be disturbed in some manner. This is quite a remarkable success given that our network operates on the standard imaging pipeline output without any advanced or bespoke processing to enhance the appearance of low surface brightness features beyond that of \citetalias{Walmsley2021}. We have provided a full breakdown of the predictions for each of the 14 galaxies in Table \ref{tab:ssls_galaxies} in Appendix \ref{app:ssls}. In most cases, the feature class predicted by our network matches reasonably well with the visual appearance of these galaxies. For the two cases where our network predicted no tidal features yet one is visible, we note that these particular features are very faint and fairly low significance according to Table 2 of \citet{Martinez-Delgado2023HiddenSurvey}. The ability of our network to recover most of the \citet{Martinez-Delgado2023HiddenSurvey} galaxies as tidally disturbed without the need for advanced image processing is seen as very promising.

\subsection{Grad-CAM Analysis}\label{sec:grad-CAM}

\begin{figure*}
    \includegraphics[width=0.97\textwidth]{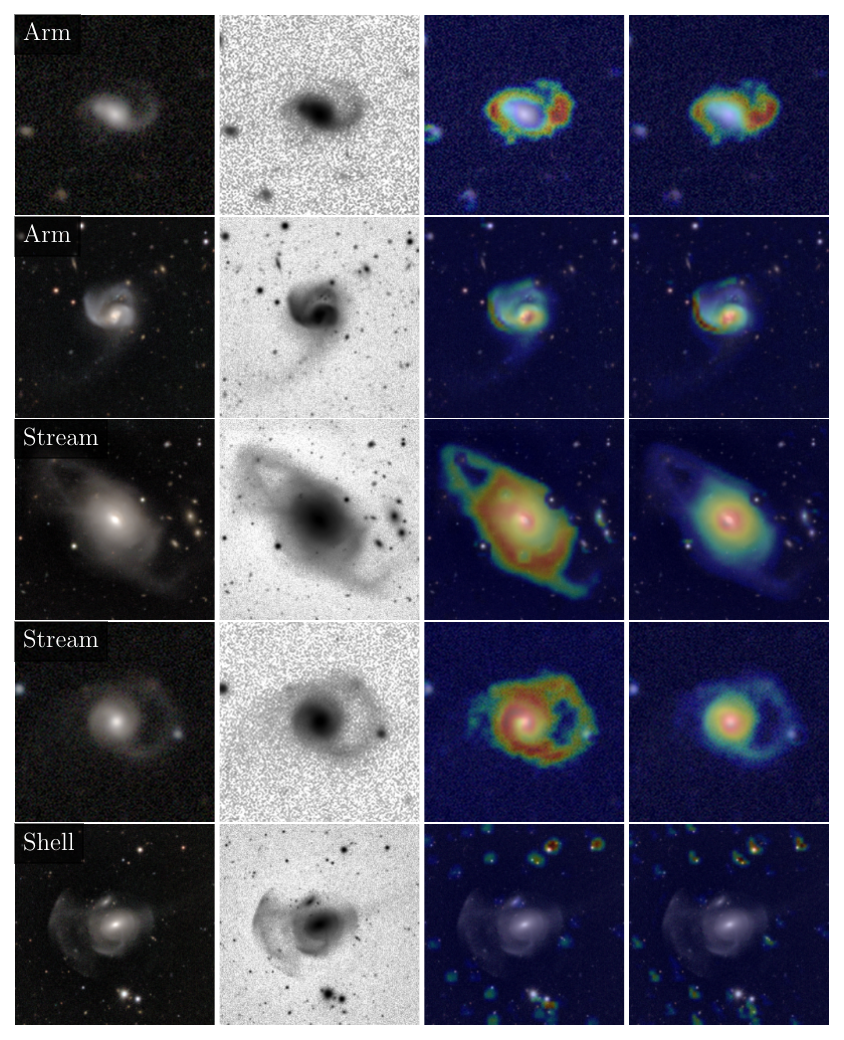}
    \caption{Outputs of a gradient-weighted class activation mapping (Grad-CAM) analysis. The analysis generated a heat map indicating where on the image the network considered important for making a particular classification (blue least to yellow/red most important). The classification being considered is written in text on each row. The leftmost column shows the image of the galaxy produced by \citetalias{Walmsley2021}. The second from the left shows the image with the low surface brightness regions highlighted according to the novel algorithm (Equation \ref{eqn:new_arcsinh_scaling}) with the parameters $M=\max({\text{pixel values}})$, $m=\text{med}({\text{background}}) - \text{stdev(background)}$, $s=4$, and $p=0.5$, where the background was estimated using a sigma clipping with a $3\sigma$ clipping limit. The two right-hand columns show the output of the Grad-CAM analysis for the two selected best-performing models: the model that reached the lowest validation loss during training and that with the greatest testing AUC score.}
    \label{fig:examplesofgradcam}
\end{figure*}

\begin{figure*}
    \includegraphics[width=0.97\textwidth]{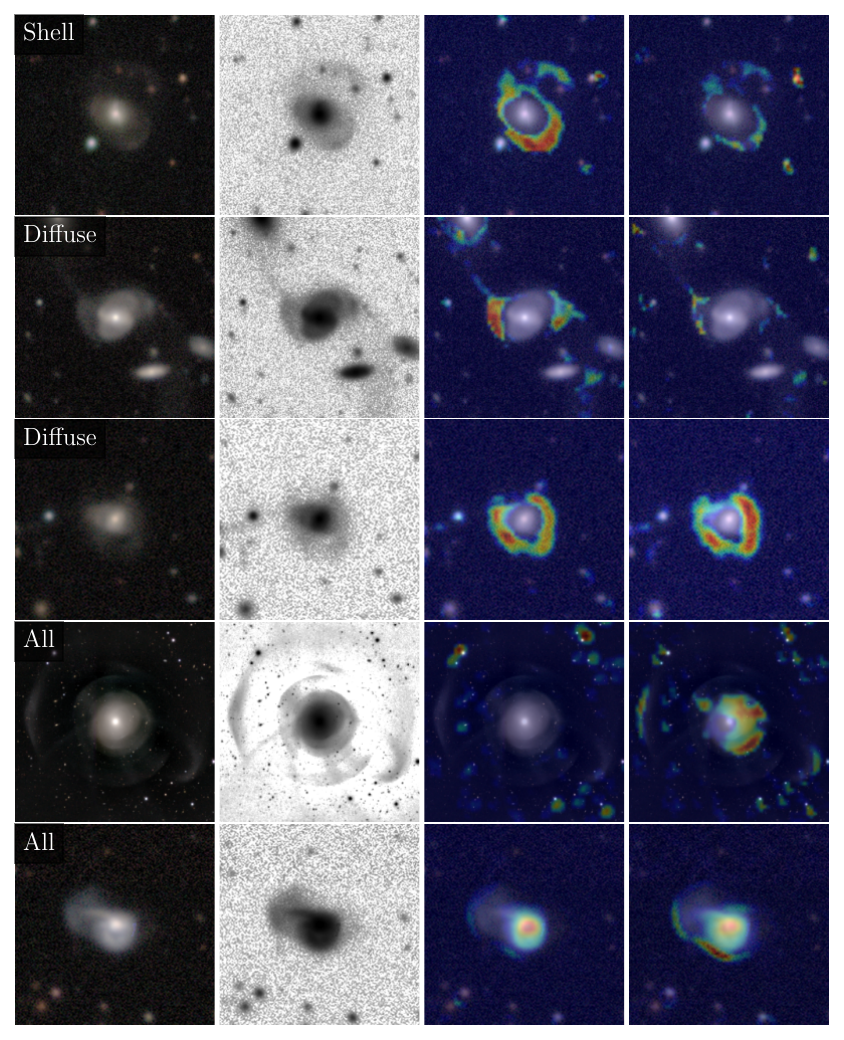}
    \contcaption{}
\end{figure*}

CNNs are prone to suffering from shortcut learning, where they can learn spurious connections between parts of the data and the desired output \citep[see, e.g.][]{Geirhos2020ShortcutNetworks}. This problem is an important issue, particularly for the generalisation of the network to unseen data, and verifying that this has not impacted the network should be a crucial part of any model development process. It is, therefore, vital to establish whether or not the network does indeed use the correct information to produce a prediction. In the context of this work, we must verify that for a given image, the network uses the galaxy and surrounding tidal features to adopt the prediction for that class.

We verified that the network was classifying the images as intended by performing a Gradient-weighted Class Activation Mapping \citep[Grad-CAM;][]{GradCAMref} analysis. In general, the analysis determines the gradient of a class score with respect to the feature map (the latent space representation) in the final convolutional layer. Then, it averages over this to assign an importance value to a particular neuron. The final heat map showing the regions of importance is generated by taking a linear combination of the importance values.

We selected the two best-performing models to analyse out of the ten attempts and five folds. These were the model with the lowest final validation loss and the model with the highest AUC score. Figure \ref{fig:examplesofgradcam} provides some examples of the original images of the galaxies (leftmost column) and a version of the image scaled using Equation \ref{eqn:new_arcsinh_scaling} (middle-left) alongside the Grad-CAM heat maps for the two models tested (two right-hand columns). Red regions on the heat map indicate areas of high importance, yellow and green lesser, and blue with little importance. 

Figure \ref{fig:examplesofgradcam} demonstrates, reassuringly, that the model identified the disturbed regions in the manner intended in almost all cases. In the case of \textit{shell} galaxies, there is some evidence that the network also identifies background galaxies and stars as regions of importance. This effect might attributed to the outer regions of these objects appearing fuzzy and faint due to limited resolution and seeing, thus resembling a \textit{diffuse} feature. Masking background sources and bright stars -- which we have not done in the present analysis -- may remove this confusion but it would come at the cost of significantly increasing the preprocessing required, which would add time and effort. Indeed, masking large samples would need to be done in an automated fashion, and there is a risk that such a procedure may perturb the appearance of the tidal features themselves. Still, we emphasise the need to strike a balance between masking unimportant background sources and maintaining the faint structure of the feature. Overall, we are very satisfied that the network appears to identify tidal features in the same manner that an expert would.

\section{Conclusions}\label{sec:conclusions}
Tidal features are a crucial probe of the hierarchical formation of galaxies, and these features contain a wealth of information that could advance our understanding of galaxy evolution. Forthcoming wide-field surveys, such as LSST and Euclid, will provide ample data to study these features at much greater limiting surface brightness depths than before. However, the volume of data from these surveys will vastly outpace the current practice of classifying tidal features using visual inspection. 

In this paper, we presented our attempts to counter this problem and demonstrate that it is possible to apply machine learning to classify faint tidal features. We used images of galaxies from the Dark Energy Camera Legacy Survey (DECaLS) to test if a Convolutional Neural Network (CNN) could classify different categories of tidal features. The images were compiled by \citet{Walmsley2021} and used to train an algorithm to reproduce volunteers' responses to questions designed to extract various morphological characteristics of the galaxy. We imposed several selection criteria using these predictions to generate a sample of 1,928 galaxies likely to have tidal features. 

We visually inspected each galaxy in the sample to identify what tidal features were present, placing them into the non-exclusive categories of \textit{arm}, \textit{stream}, \textit{shell}, \textit{diffuse}, and \textit{uncertain}. Of the 1,928 galaxies, 316 showed evidence of an \textit{arm} feature, 283 a \textit{stream}, 67 a \textit{shell}, and 517 evidence of a \textit{diffuse} feature. Furthermore, 219 galaxies showed more than one distinct type of feature. We note that a significant proportion of the galaxies could not reliably be said to have a tidal feature (labelled \textit{uncertain}) due to the potential to be confused with intrinsically irregular galaxies -- we excluded these from any further training set. Overall our sample represents one of the largest samples of galaxies with faint tidal features constructed to date.

Using this labelled set, we trained a CNN to predict what tidal features were present around each galaxy. In general, the performance was very good but depended on the feature in question, with the network appearing to be best performing for \textit{diffuse} features overall. At a level of 20 per cent contamination, our classifier retrieved a median $98.7\pm0.3$, $99.1\pm0.5$, $97.0\pm0.8$, and $99.4^{+0.2}_{-0.6}$ per cent of the true \textit{arm}, \textit{stream}, \textit{shell}, and \textit{diffuse} debris features respectively. These results are comparable to or better than other works in the literature that use machine learning to simply identify if any tidal feature is present \citep[i.e. binary detection,][]{Walmsley18, DominguezSanchez2023, Desmons2023DetectingLearning}. Furthermore, we investigated if our network was able to recover galaxies with tidal features in the overlap between our galaxy sample and that of the Stellar Stream Legacy Survey \citep{Martinez-Delgado2023HiddenSurvey}. All but two of the galaxies in the overlap were identified by our network to be disturbed in some manner without the need for any advanced or bespoke processing.

We then used a grad-CAM analysis to verify that the network was indeed classifying the images as expected and that there were no spurious connections in the data. Overall, our results provide a compelling demonstration of the potential of CNNs to classify different categories of tidal features around galaxies, although the problem remains a difficult one with this work by no means being a definitive solution. With deeper surveys and further improvements to the classification process, we aim to build up a large sample of galaxies with tidal features and use them to quantitatively study the galaxy assembly process.

\section*{Acknowledgements}
This is a pre-copyedited, author-produced PDF of an article accepted for publication in Monthly Notices of the Royal Astronomical Society following peer review. The version of record is available online at: \url{https://academic.oup.com/mnras/advance-article/doi/10.1093/mnras/stae2169/7760393}. For the purpose of open access, the author has applied a Creative Commons Attribution (CC BY) licence to any Author Accepted Manuscript version arising from this submission.

AJG acknowledges receipt of an STFC PhD studentship. AMNF is grateful for support from the UK STFC via grant ST/Y001281/1.

We thank Alice Desmons, Helena Domínguez Sánchez, Mike Walmsley, and Sarah Brough for their insightful and supportive discussions on the topic and for providing their data for our comparisons. Furthermore, we thank Sohan Seth for useful input on this work during its early stages.

This project used the following \textsc{python} packages: \textsc{astropy} \citep{AstropyCollaboration2022ThePackage}, \textsc{matplotlib} \citep{Hunter2007Matplotlib:Environment}, \textsc{numpy} \citep{Harris2020ArrayNumPy}, \textsc{pandas} \citep{Mckinney2010DataPython}, \textsc{photutils} \citep{Bradley2023Astropy/photutils:1.8.0}, \textsc{scikit-learn} \citep{Pedregosa2011Scikit-learn:Python}, \textsc{scipy} \citep{Virtanen2020SciPyPython}, and \textsc{tensorflow} \citep{Abadi2015TensorFlow:Systems}.

This project uses data from the Dark Energy Camera Legacy Survey. The Legacy Surveys consist of three individual and complementary projects: the Dark Energy Camera Legacy Survey (DECaLS; Proposal ID \#2014B-0404; PIs: David Schlegel and Arjun Dey), the Beijing-Arizona Sky Survey (BASS; NOAO Prop. ID \#2015A-0801; PIs: Zhou Xu and Xiaohui Fan), and the Mayall z-band Legacy Survey (MzLS; Prop. ID \#2016A-0453; PI: Arjun Dey). DECaLS, BASS and MzLS together include data obtained, respectively, at the Blanco telescope, Cerro Tololo Inter-American Observatory, NSF’s NOIRLab; the Bok telescope, Steward Observatory, University of Arizona; and the Mayall telescope, Kitt Peak National Observatory, NOIRLab. Pipeline processing and analyses of the data were supported by NOIRLab and the Lawrence Berkeley National Laboratory (LBNL). The Legacy Surveys project is honored to be permitted to conduct astronomical research on Iolkam Du’ag (Kitt Peak), a mountain with particular significance to the Tohono O’odham Nation.

NOIRLab is operated by the Association of Universities for Research in Astronomy (AURA) under a cooperative agreement with the National Science Foundation. LBNL is managed by the Regents of the University of California under contract to the U.S. Department of Energy.

This project used data obtained with the Dark Energy Camera (DECam), which was constructed by the Dark Energy Survey (DES) collaboration. Funding for the DES Projects has been provided by the U.S. Department of Energy, the U.S. National Science Foundation, the Ministry of Science and Education of Spain, the Science and Technology Facilities Council of the United Kingdom, the Higher Education Funding Council for England, the National Center for Supercomputing Applications at the University of Illinois at Urbana-Champaign, the Kavli Institute of Cosmological Physics at the University of Chicago, Center for Cosmology and Astro-Particle Physics at the Ohio State University, the Mitchell Institute for Fundamental Physics and Astronomy at Texas A\&M University, Financiadora de Estudos e Projetos, Fundacao Carlos Chagas Filho de Amparo, Financiadora de Estudos e Projetos, Fundacao Carlos Chagas Filho de Amparo a Pesquisa do Estado do Rio de Janeiro, Conselho Nacional de Desenvolvimento Cientifico e Tecnologico and the Ministerio da Ciencia, Tecnologia e Inovacao, the Deutsche Forschungsgemeinschaft and the Collaborating Institutions in the Dark Energy Survey. The Collaborating Institutions are Argonne National Laboratory, the University of California at Santa Cruz, the University of Cambridge, Centro de Investigaciones Energeticas, Medioambientales y Tecnologicas-Madrid, the University of Chicago, University College London, the DES-Brazil Consortium, the University of Edinburgh, the Eidgenossische Technische Hochschule (ETH) Zurich, Fermi National Accelerator Laboratory, the University of Illinois at Urbana-Champaign, the Institut de Ciencies de l’Espai (IEEC/CSIC), the Institut de Fisica d’Altes Energies, Lawrence Berkeley National Laboratory, the Ludwig Maximilians Universitat Munchen and the associated Excellence Cluster Universe, the University of Michigan, NSF’s NOIRLab, the University of Nottingham, the Ohio State University, the University of Pennsylvania, the University of Portsmouth, SLAC National Accelerator Laboratory, Stanford University, the University of Sussex, and Texas A\&M University.

BASS is a key project of the Telescope Access Program (TAP), which has been funded by the National Astronomical Observatories of China, the Chinese Academy of Sciences (the Strategic Priority Research Program “The Emergence of Cosmological Structures” Grant \# XDB09000000), and the Special Fund for Astronomy from the Ministry of Finance. The BASS is also supported by the External Cooperation Program of Chinese Academy of Sciences (Grant \# 114A11KYSB20160057), and Chinese National Natural Science Foundation (Grant \# 12120101003, \# 11433005).

The Legacy Survey team makes use of data products from the Near-Earth Object Wide-field Infrared Survey Explorer (NEOWISE), which is a project of the Jet Propulsion Laboratory/California Institute of Technology. NEOWISE is funded by the National Aeronautics and Space Administration.

The Legacy Surveys imaging of the DESI footprint is supported by the Director, Office of Science, Office of High Energy Physics of the U.S. Department of Energy under Contract No. DE-AC02-05CH1123, by the National Energy Research Scientific Computing Center, a DOE Office of Science User Facility under the same contract; and by the U.S. National Science Foundation, Division of Astronomical Sciences under Contract No. AST-0950945 to NOAO.

\section*{Data Availability}
The Galaxy Zoo: DECaLS catalogue and images originate from \citet{Walmsley2021} and are readily available at \url{https://zenodo.org/record/4573248#.ZBsa83bP2Uk}.

Other images from the Legacy Survey are available at \url{https://www.legacysurvey.org/}.

The network code is available at \url{https://github.com/aj-gordon/decals-cnn}. All further code and data for this article will be shared upon a reasonable request to the corresponding author.



\bibliographystyle{mnras}
\bibliography{references} 



\appendix
\section{Training Metrics}\label{app:metrics}

During the training of a network, it is possible to record its performance and test for overfitting. To do this, a portion of the training data was reserved as validation data, which was used to test the network's performance at every epoch. The loss metric records how well output predictions from the network match the ground truth label. We used the binary cross-entropy loss commonly used for non-exclusive multi-label classifications. The loss \begin{equation}
    \mathscr{L} = -\frac{1}{N} \sum^{N}_{i=1} \left[t_{i}\log s_{i} + (1-t_i) \log (1-s_i) \right]
\end{equation} \citep[see, e.g.][]{Murphy2012MachinePerspective} compares the truth label, $t_i$, to the output prediction, $s_i$. At every epoch, the training and validation sets were used to evaluate the loss metric; if these begin to diverge, this can indicate that the network is overfitting to the training data.

\begin{figure}
    \includegraphics[width=\columnwidth]{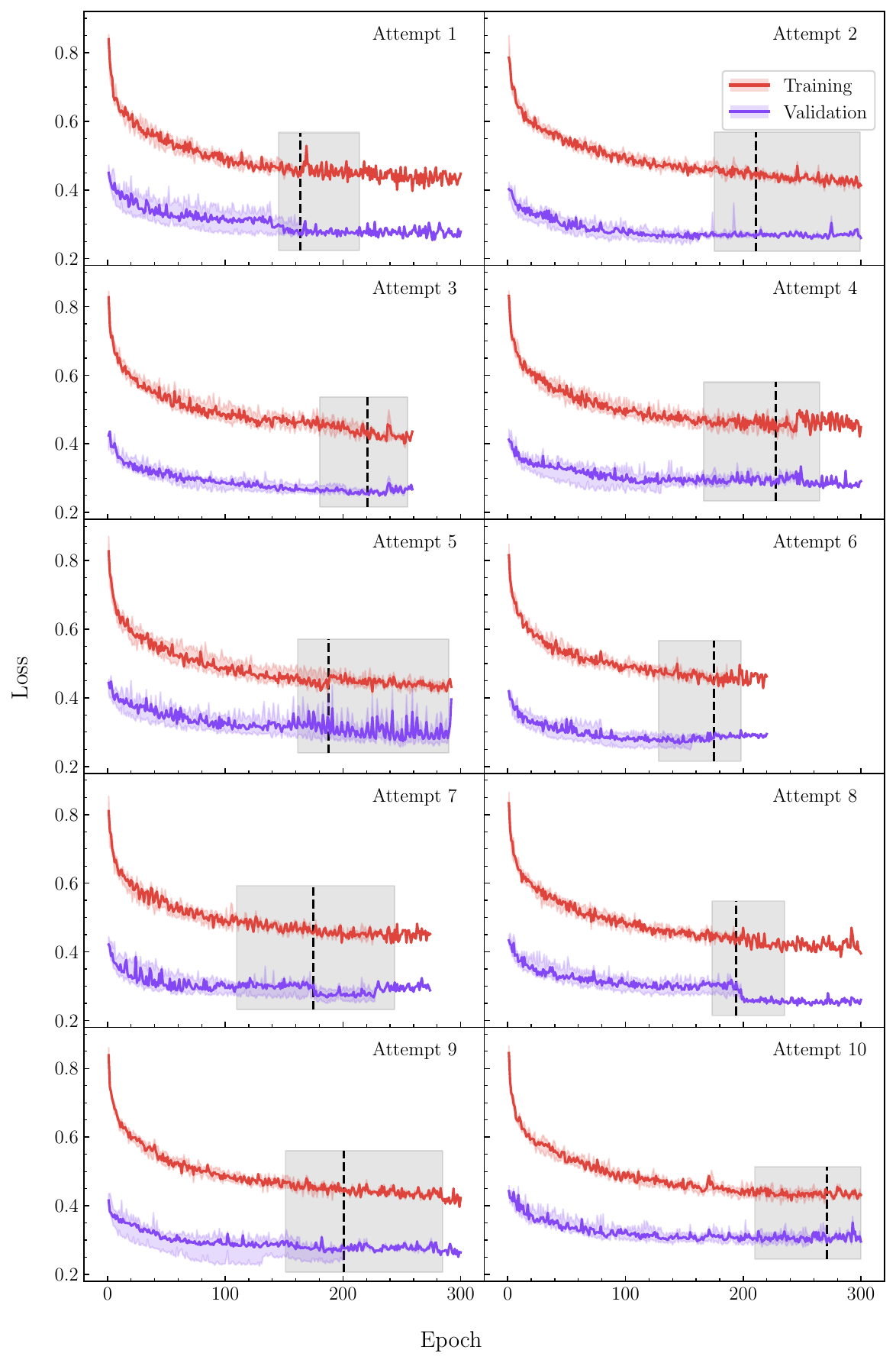}
    \caption{Training (red) and validation (blue) loss metrics evaluated at every epoch during training. Each panel shows the average across the five folds in that attempt. The black dashed line provides the median number of epochs the network was trained for, with the shaded region providing the 68.2 per cent confidence interval.}
    \label{fig:histories}
\end{figure}

Figure \ref{fig:histories} provides the training and validation losses across all epochs during training. Each panel shows the average across all of the five folds in that attempt, with the red line showing the training loss and the blue validation. The black dashed lines and shaded regions show the median epoch where training ended plus the 68.2 per cent confidence interval. In all cases, the validation loss is less than the training loss, but neither appears to diverge from the other, indicating that the network has not suffered from overfitting.

\section{Improving Classification}\label{app:improve_classifications}
During the visual inspection process, reaching a complete consensus on the type of tidal feature was often challenging due to different interpretations of the instructions and characteristics that defined each feature or due to inconsistencies in individual inspectors. We found considerable noise when considering the classifications for a particular galaxy; all three authors only agreed on every feature in 151 galaxies (15.8 per cent of the sample with tidal features). As the network was trained to recreate our labels, which we assumed to be the ground truth, we thus hypothesise that if our classifications can be improved and the noise reduced, the network performance may improve. To that end, we investigated the impact of reclassifying some galaxies, with more precise instructions and enhanced images, on the consensus between authors and how this, in turn, impacted CNN performance. 

\subsection{Reinspection}\label{sec:reinspect}

We improved the classification process by updating and clarifying the instructions in our decision tree. We included an initial question asking if any feature was present; if the inspector answered yes, they were then asked to identify what features. The choice of features was the same as in the original inspection, however we re-specified the characteristics of each feature and allowed the inspector to select whether they were confident or less confident that the feature was present.

We randomly selected 50 galaxies from our training sample to test the new classification decision tree. As before, each galaxy was given a label based on the responses from the three authors. Of the 50 galaxies, 36 per cent had no change to their label, 44 per cent had a minor change\footnote{one additional or one fewer feature.}, 12 per cent had a significant change\footnote{more than one additional or fewer features or an exchange of features.}, and 4 per cent had a critical change where the label was completely different from the original.

We found overall that the reclassification led to no significant improvement in the agreement between the authors with the same number of galaxies retaining or losing labels as those that gained labels. However, in terms of the label's confidence, we found an impact. We based the confidence of the new label on the confidence options provided, i.e. confident or less confident the feature was present. The confidence of the old label was difficult to establish but was based on the agreement between authors and whether the galaxy received votes for \textit{uncertain}. Overall, fewer labels increased confidence than those that suffered a drop in confidence.

\subsection{Retraining}\label{sec:retrain}

\begin{figure}
    \includegraphics[width=\columnwidth]{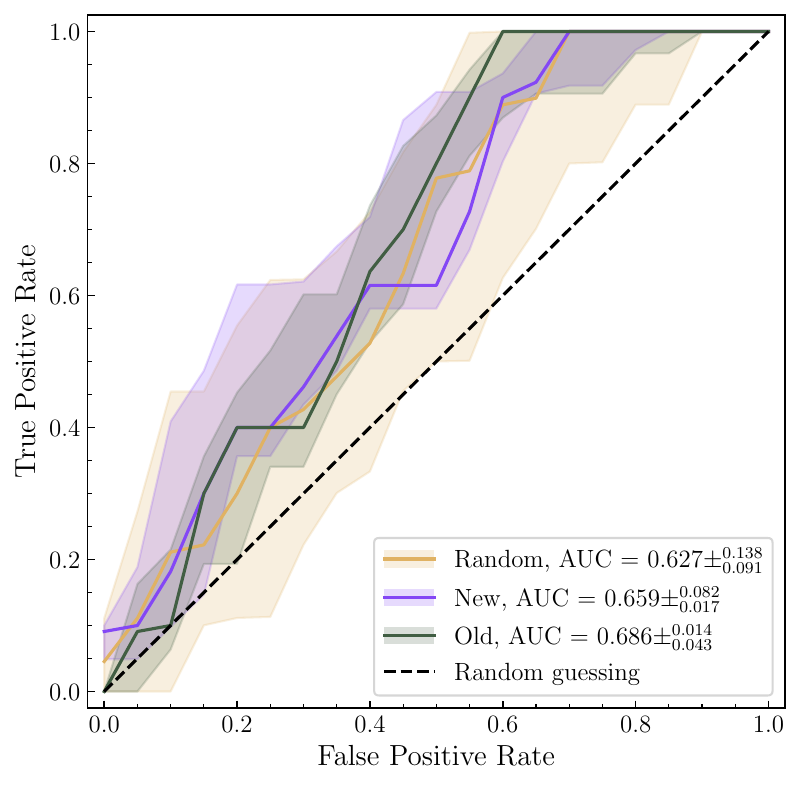}
    \caption{ROC curves from retraining the classifier on the new classifications. Each of the models in this plot was trained using $\sim$30 galaxies with the rest left for testing, the same ratios as when the classifier was trained using the full sample. The green curve provides the median of 5 trial runs of training the classifier on the original classifications of the reclassified galaxies, and the blue provides the same using the new classifications. Finally, the yellow curve is the median of classifiers trained using 20 randomly sampled sets of the full original classifications of galaxies to serve as a baseline for improved performance. Shaded regions provide the 68.2 per cent confidence interval.}
    \label{fig:reclassifications_roc}
\end{figure}

We tested the impact that the reclassifications had on the performance of the CNN. For the original and new classifications of the 50 galaxies, we trained five trial runs of the network and compared the median ROC curves for these two sets. For this experiment, rather than generate folds, we randomly selected 20 per cent of the data to be testing data at the point of loading into the network. The network's performance was then evaluated on this testing data set and averaged over the five trials. As this is such a small number of galaxies, we expect the performance to be relatively poor, so we also randomly sampled 20 groups of 50 galaxies, using the original classifications, to have a baseline comparison to the network performance at that level. Figure \ref{fig:reclassifications_roc} shows the median ROC curves for the original classifications, the new classifications, and the random sampling with the respective 68.2 per cent confidence interval shaded. We see evidence of a slight deterioration in the ROC curve based on moving from the original classifications to the new classifications. However, the median AUC scores are within $1\sigma$ of each other and there is no improvement from the randomly sampled baseline, with both the new and original AUC scores occurring well within $1\sigma$ of the baseline.

Instead of improving classifications, we suggest that other strategies, such as masking background sources, further hyper-parameter optimisation, deeper data, or transfer learning pre-trained networks, may be more effective at improving the performance of the network. Given the satisfactory performance of the network trained on the entire sample of original classifications (Section \ref{sec:multi_results}) and the time commitment to reclassifying all 956 galaxies, we decided not to continue with reclassifying all of the galaxies.

\section{Predictions of the Stellar Stream Legacy Survey Galaxies}\label{app:ssls}

Table \ref{tab:ssls_galaxies} presents the median predictions across each of our classes of tidal features for the galaxies introduced by \citet{Martinez-Delgado2023HiddenSurvey}. The predictions were generated by providing the network with the images of the galaxies and then averaging over the outputs across all of the models trained (see Section \ref{sec:multi_results} for further details). Included also are the labels assigned by the network according to the predictive threshold, $P_{\text{opt}}$, for each category introduced in Table \ref{tab:tpr_values}.

\begin{table*}
    \centering
    \caption{Predictions from our network for the galaxies introduced by the Stellar Stream Legacy Survey by \citet{Martinez-Delgado2023HiddenSurvey}.}
    \label{tab:ssls_galaxies}
    \begin{tabular}{l l c c c c c c c c}
        \hline
        \multirow{2}{*}{SSLS Galaxy} & \multirow{2}{*}{IAU Name} & \multirow{2}{*}{$z$} & \multirow{2}{*}{$\alpha (\degr)$} & \multirow{2}{*}{$\delta (\degr)$} & \multicolumn{4}{c}{Predicted Values} & \multirow{2}{*}{Label} \\
         & & & & & \textit{Arm} & \textit{Stream} & \textit{Shell} & \textit{Diffuse} &  \\
        \hline
        CGCG049-065 & J151353.24+042740.4 & 0.03747 & 228.47186 & 4.46121 & 0.3893 & 0.2999 & 0.0197 & 0.4103 & \textit{Stream} + \textit{Diffuse} \\
        IC0174 & J015616.09+034543.0 & 0.01737 & 29.06701 & 3.76187 & 0.0272 & 0.1707 & 0.7322 & 0.4667 & \textit{Shell} + \textit{Diffuse} \\
        MCG-01-06-043 & J020548.80-051739.9 & 0.01714 & 31.45256 & -5.29384 & 0.3473 & 0.2851 & 0.0441 & 0.4056 & \textit{Diffuse} \\
        NGC0095 & J002213.53+102929.6 & 0.01841 & 5.55642 & 10.49159 & 0.2824 & 0.2158 & 0.0745 & 0.3941 & \textit{Diffuse} \\
        NGC0259 & J004803.29-024631.6 & 0.01312 & 12.01367 & -2.77528 & 0.2935 & 0.1545 & 0.0023 & 0.2594 & None \\
        NGC0577 & J013040.69-015940.0 & 0.01980 & 22.66963 & -1.99437 & 0.0232 & 0.0966 & 0.7590 & 0.3595 & \textit{Shell} + \textit{Diffuse} \\
        NGC0681 & J014910.84-102535.3 & 0.00582 & 27.29506 & -10.42634 & 0.0003 & 0.0173 & 0.9527 & 0.1852 & \textit{Shell} \\
        NGC0788 & J020106.40-064855.9 & 0.01360 & 30.27693 & -6.81587 & 0.0006 & 0.0316 & 0.9474 & 0.2971 & \textit{Shell} \\
        NGC1309 & J032206.49-152400.0 & 0.00712 & 50.52731 & -15.39994 & 0.4352 & 0.1626 & 0.0071 & 0.3584 & \textit{Arm} + \textit{Diffuse} \\
        NGC3131 & J100836.40+181352.4 & 0.01702 & 152.15163 & 18.23125 & 0.1332 & 0.2214 & 0.4323 & 0.4056 & \textit{Diffuse} \\
        NGC3451 & J105420.90+271423.0 & 0.00444 & 163.58691 & 27.23966 & 0.4831 & 0.2488 & 0.0048 & 0.4020 & \textit{Arm} + \textit{Diffuse} \\
        NGC3689 & J112810.99+253940.0 & 0.00914 & 172.04591 & 25.66114 & 0.1023 & 0.1632 & 0.0487 & 0.3007 & None \\
        NGC4385 & J122542.90+003416.9 & 0.00714 & 186.42838 & 0.57258 & 0.1875 & 0.1698 & 0.0367 & 0.3777 & \textit{Diffuse} \\
        NGC4390 & J122550.90+102730.9 & 0.00363 & 186.46111 & 10.45903 & 0.4251 & 0.1714 & 0.0056 & 0.3784 & \textit{Arm} + \textit{Diffuse} \\
        ESO413-020 & & \multicolumn{8}{c}{Not imaged in DECaLS DR5}\\
        NGC0175 & & \multicolumn{8}{c}{Not imaged in DECaLS DR5}\\
        NGC5971 & & \multicolumn{8}{c}{Not imaged in DECaLS DR5}\\
        UGC04132 & & \multicolumn{8}{c}{Not imaged in DECaLS DR5}\\
        UGC06397 & & \multicolumn{8}{c}{Not imaged in DECaLS DR5}\\
        UGC08717 & & \multicolumn{8}{c}{Not imaged in DECaLS DR5}\\
        \multicolumn{2}{l}{2MASXJ12284541-0838329} & \multicolumn{8}{c}{Not imaged in SDSS}\\
        IC0160 & & \multicolumn{8}{c}{Not imaged in SDSS}\\
        IC0169 & & \multicolumn{8}{c}{Not imaged in SDSS}\\
        NGC1076 & & \multicolumn{8}{c}{Not imaged in SDSS}\\
        \hline
    \end{tabular}
\end{table*}

\bsp	
\label{lastpage}
\end{document}